\documentclass[12pt,preprint]{aastex}

\newcommand{\teff}{$T_{\rm eff} \;$}
\newcommand{\tef}{$T_{\rm eff}$}
\newcommand{\gtaus}{$\gamma \; \mathrm{Tau} \;$}
\newcommand{\dtaus}{$\delta \; \mathrm{Tau} \;$}
\newcommand{\etaus}{$\epsilon \; \mathrm{Tau} \;$}
\newcommand{\gtau}{$\gamma \; \mathrm{Tau}$}
\newcommand{\dtau}{$\delta \; \mathrm{Tau}$}
\newcommand{\etau}{$\epsilon \; \mathrm{Tau}$}
\newcommand{\cara}{$^{12}\mathrm{C}$}
\newcommand{\caras}{$^{12}\mathrm{C} \;$}
\newcommand{\carb}{$^{13}\mathrm{C}$}
\newcommand{\carbs}{$^{13}\mathrm{C} \;$}

\newcommand{\nits}{$^{14}\mathrm{N} \;$}
\newcommand{\oxy}{$^{16}\mathrm{O}$}
\newcommand{\oxys}{$^{16}\mathrm{O} \;$}

\begin{document}

\title{HYADES OXYGEN ABUNDANCES FROM THE $\lambda 6300$ [\ion{O}{1}] LINE: THE 
GIANT-DWARF OXYGEN DISCREPANCY REVISITED\altaffilmark{1,2,3}}

\altaffiltext{1}{Based on observations collected at the European Southern
Observatory, Paranal, Chile, program 70.C-0477.}
\altaffiltext{2}{This paper includes data taken with the Harlan J. Smith 2.7-m 
and the Otto Struve 2.1-m telescopes at The McDonald Observatory of the 
University of Texas at Austin.}
\altaffiltext{3}{Some of the data presented herein were obtained at the W.M. 
Keck Observatory, which is operated as a scientific partnership among the 
California Institute of Technology, the University of California and the 
National Aeronautics and Space Administration. The Observatory was made 
possible by the generous financial support of the W.M. Keck Foundation.}

\author{Simon C. Schuler\altaffilmark{4}, Artie P. Hatzes\altaffilmark{5},
Jeremy R. King\altaffilmark{4}, Martin K\"{u}rster\altaffilmark{6}, AND 
Lih-Sin The\altaffilmark{4}}
\affil{
   \altaffiltext{4}{Department of Physics and Astronomy, Clemson University, 118
   Kinard Laboratory, Clemson, SC, 29634; sschule@ces.clemson.edu,
   jking2@ces.clemson.edu, tlihsin@clemson.edu}
   \altaffiltext{5}{Th\"{u}ringer Landessternwarte Tautenburg, Sternwarte 5, 
   D-07778 Tautenburg, Germany; artie@tls-tautenburg.de}
   \altaffiltext{6}{Max-Planck-Institut f{\"u}r Astronomie, K{\"o}nigstuhl 17, 
   D-69117 Heidelberg, Germany; kuerster@mpia-hd.mpg.de}
   }

\begin{abstract}
We present the results of our abundance analysis of Fe, Ni, and O in 
high S/N, high-resolution VLT/UVES and McDonald/2dcoud{\'e} spectra of nine 
dwarfs and three giants in the Hyades open cluster.  The difference in Fe 
abundances derived from \ion{Fe}{2} and \ion{Fe}{1} lines ([\ion{Fe}{2}/H] -- 
[\ion{Fe}{1}/H]) and \ion{Ni}{1} abundances derived from moderately 
high-excitation ($\chi \approx 4.20 \; \mathrm{eV}$) lines are found to 
increase with decreasing \teff for the dwarfs.  Both of these findings are in 
concordance with previous results of over-excitation/ionization in cool young 
dwarfs. Oxygen abundances are derived from the $\lambda 6300$ [\ion{O}{1}] 
line, with careful attention given to the \ion{Ni}{1} blend.  The dwarf O 
abundances are in star-to-star agreement within uncertainties, but the 
abundances of the three coolest dwarfs ($4573 \leq T_{\mathrm{eff}} \leq 4834 
\; \mathrm{K}$) evince an increase with decreasing \tef.  Possible causes for 
the apparent trend are considered, including the effects of over-dissociation 
of O-containing molecules.  O abundances are derived from the near-UV $\lambda
3167$ OH line in high-quality Keck/HIRES spectra, and no such effects are 
found-- indeed, the OH-based abundances show an increase with decreasing \tef, 
leaving the nature and reality of the cool dwarf [\ion{O}{1}]-based O trend 
uncertain.  The mean relative O abundance of the six warmest dwarfs ($5075 \leq 
T_{\mathrm{eff}} \leq 5978 \; \mathrm{K}$) is $\mathrm{[O/H]} = +0.14 \pm 
0.02$, and we find a mean abundance of $\mathrm{[O/H]} = +0.08 \pm 0.02$ for 
the giants.  Thus, our updated analysis of the $\lambda 6300$ [\ion{O}{1}] line 
does not confirm the Hyades giant-dwarf oxygen discrepancy initially reported 
by \citet{1996AJ....112.2650K}, suggesting the discrepancy was a consequence of 
analysis-related systematic errors.  LTE Oxygen abundances from the near-IR, 
high-excitation \ion{O}{1} triplet are also derived for the giants, and the 
resulting abundances are approximately 0.28 dex higher than those derived from 
the [\ion{O}{1}] line, in agreement with NLTE predictions.  NLTE corrections 
from the literature are applied to the giant triplet abundances; the resulting 
mean abundance is $\mathrm{[O/H]} = +0.17 \pm 0.02$, in decent concordance with 
the giant and dwarf [\ion{O}{1}] abundances.  Finally, Hyades giant and dwarf O 
abundances as derived from the $\lambda 6300$ [\ion{O}{1}] line and 
high-excitation triplet, as well as dwarf O abundances derived from the near-UV 
$\lambda 3167$ OH line, are compared, and a mean cluster O abundance of 
$\mathrm{[O/H]} = +0.12 \pm 0.02$ is achieved and represents the best 
estimate of the Hyades O abundance.
\end{abstract}


\section{INTRODUCTION}
Oxygen is the most abundant element in the Galaxy after H and He.  The 
predominant nucleosynthesis site of \oxy, the most abundant O
isotope ($99.8\%$ of O in the Solar System), is He burning at the cores of 
massive stars \citep{2003hich.book.....C}.  The O is subsequently dispursed 
into the interstellar medium (ISM) when the progenitor massive stars experience 
their death throes as Type II supernovae (Woosley \& Weaver 1995).  Thus, 
mapping O abundances in various Galactic environments, such as \ion{H}{2} 
regions (e.g., Esteban et al. 2005), planetary nebulae (e.g., P{\'e}quignot, \& 
Tsamis 2005), presolar grains (e.g., Nittler et al. 1997), and stars 
(e.g., Wheeler, Sneden, \& Truran 1989; Bensby, Feltzing, \& Lundstr\"{o}m 
2004), is an irreplaceable tool in tracing the chemical evolution of the Galaxy 
and constraining supernovae rates.  Determining stellar O abundances may 
make the most crucial contribution to this endeavor.  Significant O 
overabundances relative to Fe observed in metal-poor stars is seen as a direct 
consequence of the rapid O enrichment of the ISM in the early Galaxy, resulting 
from Type II supernovae, compared to the slower enrichment of Fe, which has 
its primary nucleosynthesis site in Type Ia supernovae (Kobayashi et al. 1998; 
Matteucci \& Greggio 1986).  A clear elucidation of the [O/Fe]\footnotemark[7] 
versus Fe abundance trend is a critical component of understanding Galactic 
evolution \citep{1989ARA&A..27..279W}.

\footnotetext[7]{We use the usual bracket notation to denote abundances relative
to solar values, e.g., $\mathrm{[O/H]} = \log \{N(O)/N(H)\}_{\star} - \log
\{N(O)/N(H)\}_{\sun}$, where $\log N(H) = 12.0$.}

Deriving stellar O abundances can also provide important tests of stellar
nucleosynthesis and mixing in evolved stars (e.g., Vanture \& Wallerstein 
1999).  Stellar evolution models make precise predictions of the variations in 
surface chemical compositions due to mixing of nuclear processed material from
the core as stars evolve off the main sequence (MS; Boothroyd \& Sackman 
1999).  Known as the ``first dredge-up,'' this mixing episode is not predicted 
to alter the composition of \oxys in the atmospheres of low-mass ($\la 2.5 \;
\mathrm{M_{\sun}}$) stars.  In the present study, we rivisit the Hyades 
giant-dwarf oxygen discrepancy described by King \& Hiltgen (1996; 
henceforth KH96), who determined O abundances from the $\lambda 6300$ 
[\ion{O}{1}] line for two dwarfs and three red giants in the Hyades ($625 \; 
\pm 50 \; \mathrm{Myr}$; Perryman et al. 1998) open cluster; the O abundances 
of the giants were found to be 0.23 dex lower than those of the dwarfs.  This 
result was unexpected, and  KH96 discussed two possible sources of the 
discrepancy.  The first possibility is that the difference is real and due to 
the dredging of O-depleted material from core regions that have been mixed with 
ON-processed nuclei in the giants.  The second possibility discussed by KH96 is 
that the difference is a result of systematic errors in their analysis that may 
be related to the ability of the $\lambda 6300$ [\ion{O}{1}] line to deliver 
accurate O abundances.  Either one of these scenarios, if confirmed, could 
potentially have a considerable effect on our understanding of the evolution of 
O in the Galaxy, and we explore them more fully in the next two subsections.

\subsection{Hyades Giants and the First Dredge-Up}
The extent of the changes in surface compositions due to the first dredge-up 
are dependent on both mass and metallicity \citep{1965ApJ...142.1447I}, so 
before predictions for the Hyades giants can be compared to observations, an
estimation of their masses must be obtained (the metallicity of the Hyades is 
well-determined, e.g. Paulson, Sneden, \& Cochran 2004).  There are four 
giants in the Hyades cluster- \gtau, \dtau, \etau, and $\theta^1 \;$ Tau- all 
of which apparently reside at the same location on the cluster HR diagram and 
are of the same evolutionary state (Perryman et al. 1998; de Bruijne, 
Hoogerwerf, \& de Zeeuw 2001).  Data from the Hipparcos astrometry satellite, 
along with a 631 Myr isochrone of \citet{2000A&AS..141..371G}, have been used 
to show that the Hyades giants fall on the cluster red clump 
\citep{2001A&A...367..111D}.  Using data from \citet{1998A&A...331...81P}, de 
Bruijne et al. report masses ($M/M_{\sun}$) of 2.32, 2.30, 2.32, and 2.32 for 
\gtau, \dtau, \etau, and $\theta^1 \;$ Tau, respectively, with a common 
uncertainty of 0.10.  Fortuitously, $\theta^2 \;$ Tau, a star located at the 
turnoff in the cluster's HR diagram and a proper-motion companion to $\theta^1
\;$ Tau, is a spectroscopic binary.  The orbital solutions for the $\theta^2 
\;$ Tau system have been determined by Tomkin, Pan, \& McCarthy (1995), who 
find a mass of $M/M_{\sun} = 2.1 \; \pm 0.3$ for the primary component.  de 
Bruijne et al report a mass of $M/M_{\sun} = 2.37 \; \pm 0.10$ for the primary 
component of $\theta^2$ Tau, in good agreement with the Tomkin et al. result.  
\citet{1989ApJ...347..835G} determined a similar value for the turnoff mass 
($M/M_{\sun} = 2.2 \; \pm 0.12$) for the Hyades by fitting the cluster 
color-magnitude diagram (CMD) with the isochrones of 
\citet{1985ApJS...58..711V}.  The combined estimates of the turnoff and red 
clump masses present an evolutionary picture that is consistent with
expectations, i.e., stars at the turnoff are slightly less massive than those
that have evolved to the red giant branch (RGB).  So it seems the mass estimates
of $M/M_{\sun} \sim 2.3$ for the Hyades giants reported by 
\citet{2001A&A...367..111D} are plausible and are accepted here.

We have employed the {\it Clemson-American University of Beirut Stellar 
Evolution Code} to model the evolution and expected changes in the surface 
composition of the Hyades giants; the code is fully described in The, El Eid, 
\& Meyer (2000).  For the present study, a mass of $M/M_{\sun} = 2.5$, a 
metallicity of $Z = 0.025$ ($\mathrm{[m/H]} \approx +0.10$), the Schwarzchild 
criterion for convection, and a mixing length parameter, $\alpha$ ($\alpha = 
l/H_{\mathrm{p}}$, where $l$ is the convective scale length and 
$H_{\mathrm{p}}$ is the pressure scale height), of $2.0$ have been adopted.  
As stated above, the amount of mixing as a result of the first dredge-up 
depends on both the mass and metallicity of the star: the core radial zones 
exposed to nuclear processed material are extended outward in more massive (due 
to higher core temperatures) and more metal-poor (due to lower opacities) stars 
\citep{1994A&A...282..811C}.  We thus consider using a model characterized by a 
mass that is slightly higher than those described above as conservative, 
because the model should overestimate the amount of mixing that has occurred in 
the Hyades giants.  The adopted metallicity is characteristic of that for the 
Hyades, which is well-known to be metal-rich compared to the Sun (see \S 3.3).  
The model was evolved approximately 800 Myr and was stopped during He shell 
burning phase; the temporal scale of our model should be sufficient for the 
Hyades giants, which have an age of about 625 Myr.  The stellar structure as a 
function of time is given in Figure 1; time is plotted on a logarithmic scale 
in terms of ($\mathrm{t}_{\mathrm{f}} - \mathrm{t}$), where 
$\mathrm{t}_{\mathrm{f}} \approx 7.9 \times 10^8 \; \mathrm{yr}$, the total 
elapsed time, and $\mathrm{t}$ is time since $\mathrm{t} = 0$.  Convective 
regions are shown as gray areas, and radiative regions are white.  The core 
remains convective for nearly 475 Myr, during which time it contracts at a near 
constant rate.  Afterwards, shell H burning commences, and the core becomes 
radiative.  This continues for a short duration until the conditions in the 
core are sufficient so that He burning begins, and the star experiences the 
first dredge-up.  At this time, nuclear processed material from the core is 
mixed with the pristine material in the atmosphere.  It can be seen in Figure 1 
that at no subsequent stage in the star's evolution does the surface convection 
zone extend deep enough to dredge further processed material.
\marginpar{Fig.~1}

Changes in the surface composition of the Hyades giants predicted by our $2.5 
\; M_{\sun}$ model can now be compared to observations.  The CNO bi-cycle is 
the main energy generation source for stars with masses greater than about 1.3 
$M_{\sun}$, and the abundances of some of the isotopes involved with the 
reactions- \cara, \carb, and \oxy- are of interest here.  The general 
expectation is that the abundances of \caras and \oxys are reduced and those of 
\carbs are enhanced in the convective core as a result of the reactions 
comprising the bi-cycle (e.g., El Eid 1994).  Changes in the abundances of 
these isotopes are manifested at the surface as a result of the first 
dredge-up.  The first reaction of the bi-cycle converts \caras into \carbs via 
the reaction 
$^{12}\mathrm{C}(p,\gamma)^{13}\mathrm{N}(\beta^{+}\nu)^{13}\mathrm{C}$ and is 
followed by the conversion of \carbs into \nits via 
$^{13}\mathrm{C}(p,\gamma)^{14}\mathrm{N}$.  The second of these reactions lags
the first due to its dependence on the \carbs abundance; the general result is a
decrease in the \caras abundance and an increase in the \carbs and \nits 
abundances.  Indeed, the surface \cara/\carbs ratio as our $2.5 \; M_{\sun}$ 
model evolves onto the MS is 90.0 and reduces to 23.6 after the first 
dredge-up, and as expected, the ratio remains constant following this initial 
reduction (Figure 2).  Spectroscopic determinations of the \cara/\carbs ratio 
for the Hyades giants are available in the literature, and our theoretical 
value is in good concordance with the observational results.  
\citet{1976ApJ...210..694T} found \cara/\carbs ratios of 19, 23, 22, and 20 for 
\gtau, \dtau, \etau, and $\theta^1 \; $ Tau, respectively, and 
\citet{1989ApJ...347..835G} reported ratios of 26.0, 24.0, 25.5, and 
27.5\footnotemark[8] for the same four stars.  The reported uncertainty of the 
Tomkin et al. (1976) values are $\pm 15\%$, and no uncertainty is given by 
Gilroy.  Again, these \cara/\carbs ratios are in good agreement with our 
theoretical value.  
\marginpar{Fig.~2}

\footnotetext[8]{The quoted ratios from Gilroy (1989) are the averages of the 
two values given in Table 4 therein.}

Another check of the plausibility of our model is the overall reduction of 
\caras after the first dredge-up.  The dominant isotopic species for C is 
\cara, making up $98.9\%$ Solar System composition, so observationally 
determined C abundances are equivalent to the model \caras abundance.  
Comparing the initial and final \caras abundances from our model, a reduction 
by a factor of 1.56 is found.  Comparison to observations is more uncertain in 
this case because of the lack of accurate C abundances for the Hyades dwarfs.  
We have thus determined C abundances for two dwarfs in our sample (see \S 
3.4.2) and compared the mean abundance to that of the giants as reported by 
KH96, which is based on previous determinations.  The reduction factor is 
approximately 1.8; again we find satisfactory agreement between our model and 
observations.  

With regards to $^{16}\mathrm{O}$, the abundance of this isotope should decline 
due to its destruction via the first reaction of the secondary component of the 
bi-cycle (ON-cycle), 
$^{16}\mathrm{O}(p,\gamma)^{17}\mathrm{F}(\beta^{+}\nu)^{17}\mathrm{O}$.  The
ON-cycle is not efficient in low mass stars due to the relatively low core 
temperatures, and a large reduction of \oxys is not expected.  Comparing the
initial \oxys surface abundance to that after the first dredge-up, our model 
predicts a decrease of less than $3.5\%$, with no alteration of the \oxy 
abundance seen prior or subsequent to the first dredge-up (Figure 2).  If we 
consider the O abundances of KH96, a comparison to our theoretical prediction 
can be made.  The mean O abundance for the giants is approximately $41\%$ lower 
than that for the dwarfs, a value that far exceeds the quoted uncertainty in 
the KH96 results.  

The changes in surface compositions due to the first dredge-up predicted by our 
$2.5 \; M_{\sun}$ are typical of similar calculations.  Although specific 
composition alterations are dependent on mass and metallicity, the general 
results include a decrease in the \cara/\carbs ratio from $\sim 100-90$ to 
$\sim 20-25$, a reduction of about 1.5-2 in the $^{12}\mathrm{C}$ abundance, 
and a negligible change in the $^{16}\mathrm{O}$ abundance (e.g., Sweigart, 
Greggio, \& Renzini 1989; Charbonnel 1994; Boothroyd \& Sackmann 1999).  There
are observational data suggesting some RGB stars experience an extra-mixing 
episode.  The primary evidence for this is lower than expected \cara/\carbs 
ratios, with values decreasing as low as 6  (e.g., Gilroy \& Brown 1991; 
Gratton et al. 2000).  However, this extra-mixing seems to take place at or 
subsequent to the RGB bump and is related to rotation (Charbonnel 2004; Weiss 
\& Charbonnel 2004).  Interestingly, the surface $^{16}\mathrm{O}$ abundance of 
RGB stars that have experienced the extra-mixing is still not altered 
significantly (Charbonnel 2004; Gratton et al. 2000).  One final supporting 
piece of evidence that the Hyades giants have not experienced non-standard 
mixing comes from the study of \citet{1998ApJ...499..871D}, who derived the 
abundance of B from HST/GHRS spectra for two giants and one turnoff star in the 
Hyades.  Boron is a fragile element which is destroyed at a moderately low
temperature of $5.0 \times 10^6 \; \mathrm{K}$.  Duncan et al. found the B
abundances to agree nicely with predictions from models that include standard
mixing only.  We thus conclude, like KH96, that post-MS mixing does not appear 
to be a plausible explanation for the discrepancy in the O abundances of the 
Hyades giants and dwarfs.  

\subsection{Spectroscopic Oxygen Abundances}
Those who undertake spectroscopic determinations of stellar O abundances are
fully aware of the intricacies of the handful of spectral features available 
for this purpose.  Abundances derived from molecular OH bands in the 
IR and near-UV are highly-sensitive to the adopted effective temperature 
(\teff), and many lines (especially in the near-UV) suffer from blends (e.g., 
Asplund \& Garc{\'{\i}}a P{\' e}rez 2001).  The high-excitation, near-IR 
triplet at 7771 - 7775 {\AA} (henceforth referred to as the triplet) is 
susceptible to non-LTE (NLTE) effects in the atmospheres of giants (e.g., 
Kiselman 1991; Takeda 2003) and warm dwarfs ($T_{\mathrm{eff}} \ga 6100 \; 
\mathrm{K}$), and recent studies of the triplet in cool open cluster stars 
suggest it fails as an accurate O abundance indicator for dwarfs with 
$T_{\mathrm{eff}} \la 5500 \; \mathrm{K}$ (e.g., Schuler et al. 2005).  The 
last set of lines, resulting from forbidden transitions manifested at 6300.30 
and 6363.78 {\AA}, are weak in the spectra of solar-type dwarfs 
($\la 5.5 \; \mathrm{m\AA}$ for the Sun) and are blended (Asplund et al. 2004).

It is the forbidden lines, particularly the $\lambda 6300$ feature, that are 
believed to be the most well-suited for abundance derivations, because they are 
not particularly sensitive to either atmospheric model parameters or NLTE 
effects.  However, proper account for the blending lines is required if 
accurate abundances are going to be obtained.  Fastening the discussion on the 
$\lambda 6300$ line, which was used by KH96 and in the present study, the 
presence of a blending \ion{Ni}{1} line is now firmly established (e.g., 
Johansson et al. 2003).  The identification of the blend as a Ni line was first 
made by \citet{1978MNRAS.182..249L}, and it took almost two decades before the 
blend was subjected to fine analyses and before more accurate estimates of its 
oscillator strength ($gf$-value) were obtained (e.g., Allende Prieto, Lambert, 
\& Asplund 2001).  These studies were subsequent to KH96, who used a $gf$-value 
for the Ni blend based on an inverted analysis using the solar Ni abundance of 
\citet{1989GeCoA..53..197A} and equivalent width (EW) estimates of 
\citet{1982A&A...115..145K}.  Their value, $\log gf = -3.00$, is about 1 dex 
lower than modern values (see \S 3.4).  

The updated $gf$-value of the Ni blend has provided the primary motivation to
revisit the Hyades giant-dwarf O discrepancy of KH96.  Because of the potential
implications of a real O discrepancy, we have made a concentrated effort to
improve on the KH96 analysis.  Besides the use of updated atomic parameters, 
our analysis includes original, high-quality spectra of nine dwarfs and three 
giants in the Hyades cluster.  Oxygen abundances derived from the weak 
$\lambda 6300$ [\ion{O}{1}] line are affected by the adopted Ni and C (because 
of the CO molecule) abundances, so Ni abundances have been derived for each 
star and C for two dwarfs.  One-dimensional model atmospheres have been 
interpolated from grids utilizing different values of the mixing length 
parameter (see \S 3.2), which may be important when making comparisons between 
dwarfs and giants (Demarque, Guenther, \& Green 1992).  The discussion of our 
analysis and results is presented in the following manner: details of the 
observations and data reduction are given in \S 2; in \S 3, the analysis 
techniques and results are presented; a discussion of the results follows in 
\S 4; and the paper concludes with a summary in \S 5.

\section{OBSERVATIONS AND DATA REDUCTION}
Multiple spectra of seven Hyades dwarfs were obtained in service mode (program 
70.C-0477) with the VLT Kueyen (UT2) telescope and UVES high resolution 
spectrograph in 2002 October and 2003 January as part of an ongoing radial 
velocity survey.  The data used here are the template spectra, i.e., were taken 
without an iodine cell.  The instrumental setup included the red arm and CD3 
cross disperser ($600 \; \mathrm{g} \; \mathrm{mm}^{-1}$), resulting in a 
wavelength coverage of 4920-7070 {\AA}, with a side-by-side mosaic of two $2048 
\times 4096$ EEV CCDs.  Preslit optics included the \#3 image slicer, providing 
five slices per order, and a slit width of $0.3''$ yielded a resolution of $R 
\approx 110,000$.  Integration times ranged from 120 s for the brightest star 
observed (HIP 19148) to 600 s for the faintest (HIP 22654); signal-to-noise 
(S/N) ratios range from 145 - 200 per exposure.  Two additional Hyades dwarfs 
have been observed with the Harlan J. Smith 2.7-m telescope and the
``2dcoud{\'e}'' cross-dispersed echelle spectrometer at The McDonald 
Observatory on 2004 October 12.  The spectra are characterized by a resolution 
of $R \approx 60,000$ and S/N ratios of 175 and 225.  The McDonald observations 
are fully described in \citet{trip}.

The Hyades giants \gtau, \dtau, and \etaus have been observed with both the
2.7-m and the Otto Struve 2.1-m telescopes at The McDonald Observatory.  The
2.7-m observations were carried out on 2004 October 10 during the same 
observing run as the two additional dwarfs noted above.  Each giant was 
observed once for 300 s, and the resulting S/N ratios for \gtau, \dtau, and 
\etaus are 700, 500, and 640, respectively.  The 2.1-m observations utilized 
the Sandiford Cassegrain echelle spectrograph \citep{1993PASP..105..881M}, 
along with a Reticon $1200 \times 400$ CCD and a slit width of $\sim 1.0''$; a 
resolution of $R\approx 60,000$ was achieved.  Two exposures each of \gtaus and 
\dtaus were taken on 10 Sept 1994 (UT), and three exposures were taken of 
\etaus on 10 October 1994 (UT).  Integration times for \gtau, \dtau, and \etaus 
were 110, 89, and 58 s resulting in S/N ratios of 530, 485, and 475, 
respectively, for each exposure.  A log of the observations, as well as stellar 
cross-identifications, are presented in Table 1.
\marginpar{Tab.~1}

Spectra obtained with VLT/UVES were reduced utilizing the {\sf ESO-MIDAS} 
system.  The data have been subjected to bias subtraction, flatfielding,
scattered light removal, and cosmic ray removal.  The orders were then extracted
and wavelength calibrated.  Reductions of the dwarf and giant McDonald spectra 
were carried out using standard {\sf IRAF}\footnotemark[9] routines for 
echelle spectra.  Bias subtraction, flat fielding, scattered light corrections, 
extraction, and dispersion solutions were all performed.  Extracted spectra 
for the dwarfs and the giants with multiple exposures obtained during a given
observational run were co-added for the final analysis.

\footnotetext[9]{IRAF is distributed by the National Optical Astronomy 
Observatories, which are operated by the Association of Universities for 
Research in Astronomy, Inc., under cooperative agreement with the National 
Science Foundation.}

\section{ANALYSIS \& RESULTS}
LTE derivations of Fe, Ni, and [\ion{O}{1}] abundances for both the dwarfs and 
giants have been carried out with an updated version of the LTE stellar line 
analysis package {\sf MOOG} (Sneden 1973; Sneden 2004, private communication).  
Equivalent widths of the chosen Fe and Ni lines, as well as the [\ion{O}{1}] 
feature, have been measured using Gaussian profiles with the one-dimensional 
spectrum analysis package {\sf SPECTRE} \citep{1987BAAS...19.1129F}.  Solar 
equivalent widths have been measured from a high-quality ($\mathrm{S/N} \sim 
1000$ and $R \sim 60,000$) echelle spectrum of the daytime sky obtained at the 
Harlan J. Smith 2.7-m telescope in 2004 October.

\subsection{Stellar Parameters}
\subsubsection{The Dwarfs}
Stellar parameters for all of the dwarfs except HIP 22654 and HD 29159 are taken
from the recent study of \citet{trip}, where the reader will find a discussion
on the derivation process.  The stellar parameters for HIP 22654 and HD 29159, 
two stars not included in the Schuler et al. sample, were derived using the 
same method described therein.  The (\bv) colors for these two stars are
from \citet{2004ApJ...603..697Y}.

\subsubsection{The Giants}
Stellar parameters for the Hyades giants have been derived by numerous groups,
and the values from the more recent literature have been chosen here.  Effective
temperatures for \gtaus and \dtaus are form \citet{1994A&A...282..899B}, 
and that for \etaus is from \citet{1998A&AS..129..505B}.  Both of these 
studies derive \teff using the infrared flux method, and we 
estimate the $1 \sigma$ errors to be $\pm 75 \; \mathrm{K}$.  Microturbulent 
velocities ($\xi$) and surface gravities ($\log g$) for \gtaus and \etaus are 
from \citet{1999A&A...350..859S}, who derived atmospheric parameters by fitting 
synthetic spectra- synthesized using LTE model atmospheres- to observed 
high-resolution ($0.05 \; \mathrm{\AA}$ full-width at half-maximum) spectra of 
these two stars.  The microturbulent velocities of Smith (1999) are $\sim 0.70
\; \mathrm{km} \; \mathrm{s}^{-1}$ lower than those adopted by other groups 
that have derived $\xi$ by requiring zero correlation between derived 
abundances and equivalent widths of the measured Fe lines (e.g., McWilliam 
1990; Gilroy 1989).  However, the lower values are similar to those found for 
other giants at equivalent points in their evolution 
\citep{1999A&A...350..859S} and are thus adopted here.  For the sake of 
consistency, the microturbulent velocity for \dtaus has been taken to be the 
average of those for \gtaus and \etau.  It is shown below that the choice of
$\xi$ has little affect on the O abundances derived from the [\ion{O}{1}] 
feature and is not of critical importance in the present analysis.  Finally, 
the $\log g$ value for \dtaus is from KH96, who determined a physical gravity 
by using \tef, V magnitude, mass, distance modulus, and bolometric corrections 
from the literature.  KH96 estimated the $1 \sigma$ errors in the $\log g$ 
values to be 0.12 dex, while no error was given by 
\citet{1999A&A...350..859S}.  We have chosen a conservative $1 \sigma$ error of 
$\pm 0.15 \; \mathrm{dex}$ for the surface gravities adopted here.  Final 
adopted atmospheric parameters, including those for the Sun, are given in 
Table 2.
\marginpar{Tab.~2}

\subsection{Model Atmospheres}
LTE model atmospheres characterized by the metallicity of Paulson et al. (2003) 
($\mathrm{[Fe/H]} = +0.13$) and the adopted parameters for each star have been 
interpolated from two sets of ATLAS9 grids, those with convective overshoot and 
a mixing length parameter $\alpha = 1.25$\footnotemark[10] ($\alpha = 
l/H_{\mathrm{p}}$, where $l$ is the convective scale length and 
$H_{\mathrm{p}}$ is the pressure scale height) and those with no convective 
overshoot and $\alpha = 0.50$ \citep{2002A&A...392..619H}.  Schuler et al. 
(2004) recently derived O abundances from the triplet for 15 Pleiades dwarfs 
and from the [\ion{O}{1}] forbidden line for three of the dwarfs using 
interpolated models from the two grids used here, as well as ATLAS9 grids with 
no convective overshoot and $\alpha = 1.25$ \citep{1997A&A...318..841C} and 
ATLAS9 grids that have been modified to include the convective treatment of 
\citet{1996ApJ...473..550C}.  The Pleiades O abundances derived from both 
features were found to be independent of model atmosphere.  Nonetheless, the 
MLT5 models have not been used to derive metal abundances for red giants, and 
given possible differences in the giant and dwarf values of $\alpha$ 
\citep{1992AJ....103..151D}, comparing the OVER and MLT5 results is of interest 
here.

\footnotetext[10]{See http://kurucz.harvard.edu/grids.html}

\subsection{Fe \& Ni Abundances}
The abundances of Fe and Ni have been derived for both the dwarfs and the 
giants from a set of clean lines in the spectral region 5807-6842 {\AA}.  Each 
line was carefully scrutinized for known atomic and telluric contamination, and 
only lines that are free from blends at a high confidence level have been used. 
Wavelengths, lower excitation potentials, oscillator strengths ($gf$-values), 
and measured equivalent widths for each line are given in Tables 3 and 4.  
Transition probabilities are from the VALD database (Piskunov et al. 1995; 
Kupka et al. 1999; Ryabchikova et al. 1999).  Final Fe and Ni abundances are 
given relative to the Sun via a line-by-line comparison, which minimizes 
uncertainties in the final abundances due to $gf$-values.  
\marginpar{Tab.~3}
\marginpar{Tab.~4}

Iron abundances have been derived primarily to confirm previously determined 
super-solar [Fe/H] values for the Hyades cluster and to substantiate our choice 
of the atmospheric model metallicity.  Iron abundances derived from neutral and 
singlely ionized lines using the interpolated models with overshoot (OVER) and
without overshoot (MLT5) are given in Table 5, where total uncertainties in the 
final abundances are also provided.  The total error in the Fe and Ni 
abundances is the quadratic sum of the uncertainty in the mean abundances and 
the abundance uncertainties associated with \teff, $\log g$, and $\xi$.  
Abundance sensitivities to the atmospheric parameters were calculated by 
considering changes in \teff, $log g$, and $\xi$ of $\pm 150 \; \mathrm{K}$, 
$\pm 0.25 \; \mathrm{dex}$, and $\pm 0.30 \; \mathrm{km} \; \mathrm{s}^{-1}$, 
respectively. 
\marginpar{Tab.~5}

It is seen clearly in Table 5 that the individual stellar abundances of both 
Fe species, as well as of \ion{Ni}{1}, are independent of model atmosphere.  A 
similar result was found by the aforementioned analysis of the \ion{O}{1} 
triplet in the spectra of 15 Pleiades dwarfs by \citet{2004ApJ...602L.117S}.  
More interesting here is the model independence of the giant Fe and Ni 
abundances.  The value of $\alpha$ is a function of the physics used in the 
construction of a stellar model, and it has been suggested that different 
values of $\alpha$ are required for modeling MS dwarfs than for stars that have 
evolved onto the RGB \citep{1992AJ....103..151D}.  In an attempt to match the 
convective efficiency of standard mixing length theory models to that of 2D 
radiation hydrodynamics calculations, \citet{1999sstt.conf..225F} found that an 
increase in the value of $\alpha$ was required to match the 2D results for the 
Sun as it evolves off the MS.  Regardless, the value of $\alpha$ does not 
appear to be important in the derivation of metal abundances when atmospheric 
models interpolated from ATLAS9 LTE grids are used: the giant OVER and MLT5 
abundance differences for each star presented here is $\leq 0.02 \; 
\mathrm{dex}$.  Subsequent references in this report to our derived dwarf and 
giant Fe abundances will quote the OVER value.

Inspection of Table 5 also reveals that the dwarf Fe abundances as derived from 
the \ion{Fe}{2} features are generally greater than those derived from the 
\ion{Fe}{1} features.  The abundance differences ($\Delta \mathrm{Fe} = $
[\ion{Fe}{2}/H] $-$ [\ion{Fe}{1}/H]) are plotted in Figure 3, where an 
increasing discrepancy with decreasing \teff is seen for $T_{\mathrm{eff}} 
\lesssim 5400 \; \mathrm{K}$.  A similar result was reported initially by Yong 
et al. (2004; Figure 4 therein), who measured Fe abundances from both 
\ion{Fe}{2} and \ion{Fe}{1} lines in 56 Hyades dwarfs in the range $4000 
\lesssim T_{\mathrm{eff}} \lesssim 6200 \; \mathrm{K}$.  Although the $\Delta
\mathrm{Fe}$ values presented here appear to start increasing at a higher \teff 
and in a more linear fashion than those reported by Yong et al., the trends 
agree nicely within uncertainties.  In addition to the Hyades, dwarf $\Delta 
\mathrm{Fe}$ abundance trends have been reported for M34 
\citep{2003AJ....125.2085S} and the Ursa Major (UMa) moving group 
\citep{ksuma}; the sample size of these later two studies is discouragingly 
small, making intercluster comparisons difficult.  The lack of ionization
balance is not limited to open cluster dwarfs, as it has been observed in nearby
field stars, as well (Allende Prieto et al. 2004; Ram{\'i}rez 2005).  
Observations of a large (20-30) number of dwarfs spanning \teff of at least 
$4500 \leq T_{\mathrm{eff}} \leq 6000 \; \mathrm{K}$ in a handful of clusters 
with differing ages and metallicities are needed in order to address the nature 
of $\Delta \mathrm{Fe}$ trends.
\marginpar{Fig.~3}

Calculating an accurate mean cluster Fe abundance is difficult given the $\Delta
\mathrm{Fe}$ trends discussed above.  Including the \ion{Fe}{2} abundances, even
of the warm dwarfs, could artificially inflate the mean abundance.  Utilizing 
the \ion{Fe}{1} lines only, the mean Fe abundance is $\mathrm{[Fe/H]} = +0.08 
\; \pm 0.01$ (uncertainty in the mean) for the dwarfs and $\mathrm{[Fe/H]} = 
+0.16 \; \pm 0.02$ for the giants.  Both of these values are in good agreement 
with previous determinations for the Hyades, e.g., $\mathrm{[Fe/H]} = +0.12 \; 
\pm 0.03$ \citet{1985A&A...146..249C}, $0.13 \; \pm0.01$ (Paulson et al. 2003), 
and $0.10 \pm0.01$ \citep{2004AAS...204.0716T}, and thus, we consider our 
choice of 0.13 dex for the interpolated model metallicities as reasonable and 
appropriate for the Hyades [\ion{O}{1}] analysis.  We note that altering the 
adopted model [Fe/H] abundances, i.e. metallicities, by $\pm 0.05 \; 
\mathrm{dex}$ results in a modest change in the O abundance derived from the 
[\ion{O}{1}] feature, with typical differences $< 0.04 \; \mathrm{dex}$.

\subsection{Oxygen Abundances}
The {\sf blends} driver within the {\sf MOOG} software package, along with the 
measured [\ion{O}{1}] equivalent widths and interpolated model atmospheres, was 
used to derive the O abundances of the dwarfs and giants.  
This driver accounts for the blending feature- Ni $\lambda 6300.34$ in this 
case- when force fitting the abundance of the primary element- [\ion{O}{1}] 
here- to the measured equivalent width of the blended feature.  In order 
to derive a sound abundance of the primary element of a blend, accurate atomic 
parameters ($gf$-values) for both the primary and blending features, as well as 
a good guess of the blending feature's abundance, are needed.  The relative Ni 
abundance derived for each individual star was used for the atmospheric model 
input abundance for each exclusively, and the Ni abundance from 
\citet{1989GeCoA..53..197A} ($\log N\mathrm{(Ni)}=6.25$) was adopted for the 
input abundance for the Sun.  

We utilize the [\ion{O}{1}] $gf$-value from the fine $\chi^2$-analysis of 
Allende Prieto et al. (2001), $\log gf = -9.717$).  Allende Prieto et al. also 
found $\log gf = -2.31$ for the Ni $\lambda 6300.34$ blend by 
treating the Ni $gf$-value as a free parameter in their analysis.  More
recently, \citet{2003ApJ...584L.107J} has shown experimentally that the Ni 
feature is actually due to two isotopic components, $^{58} \mathrm{Ni}$ and 
$^{60} \mathrm{Ni}$, and find $\log gf = -2.11$ for the unresolved Ni feature.  
Following the recommendation of Johansson et al., \citet{2004A&A...415..155B} 
calculated the weighted $gf$-values of the individual components by assuming a 
solar isotopic ratio of 0.38 for $^{60} \mathrm{Ni}$ to $^{58} \mathrm{Ni}$; 
the resulting values are $\log gf(^{60} \mathrm{Ni}) = -2.695$ and $\log 
gf(^{58} \mathrm{Ni}) = -2.275$.  These weighted isotopic values have been used 
in our analysis.  Oxygen abundances were also derived using the single value of 
Johansson et al., and the resultant abundances differ from those derived with 
the individual isotopic values by no more than 0.01 dex for both the dwarfs and 
giants.  Equivalent widths, which are the mean values of three individual
measurements, and final O abundances for the dwarfs, giants, and
the Sun are given in Table 6.
\marginpar{Tab.~6}

\subsubsection{The Sun}
The solar [\ion{O}{1}] equivalent width as measured from our high-quality
McDonald 2.7-m spectrum of the daytime sky is $5.5 \; \mathrm{m\AA}$, the same 
value measured in our previous study of Pleiades and M34 dwarfs 
\citep{2004ApJ...602L.117S}.  KH96 synthesized the solar $\lambda 6300$ region 
and reported a computed equivalent width for the [\ion{O}{1}] feature of $5.4 \;
\mathrm{m\AA}$.  \citet{1992A&A...261..255N} conducted a study of relative O
abundances from the [\ion{O}{1}] feature in field F and G dwarfs and measured 
solar [\ion{O}{1}] equivalent widths of 5.8 and 5.6 from two separate solar 
spectra obtained with the ESO 1.4-m Coud{\'e} Auxiliary Telescope and 
Coud\'{e} Echelle Spectrometer, as well as measuring a value of 5.3 from the
Solar Flux Atlas of \citet{1984sfat.book.....K}.  Our measured value agrees 
nicely with these previous measurements, with the largest difference being 
$0.3 \; \mathrm{m\AA}$.  A $\pm0.5$ difference in our adopted solar equivalent 
width results in a $\pm 0.06$ difference in the derived solar O abundance. 

Molecular equilibrium calculations were included in the O abundance derivations,
with CO being the only molecule with any significant effect on the calculations.
An input C abundance of $\log N(\mathrm{C}) = 8.56$ \citep{1989GeCoA..53..197A} 
was used for the Sun.  The derived solar O abundances are $\log N(\mathrm{O}) = 
8.69 \; \pm 0.06$ (total internal uncertainty) and $\log N(\mathrm{O}) = 8.66 
\; \pm 0.06$ using the OVER and MLT5 models, respectively.  Again it is seen 
that the different model atmospheres produce the same abundance to within a few 
hundredths of a dex, and the following discussion will refer to the OVER
abundance.  The solar O abundance found here is 0.23 dex lower than
``traditional'' value derived utilizing the [\ion{O}{1}] $\lambda 6300$ 
feature \citep{1978MNRAS.182..249L} and 0.14 dex lower than the updated 
abundance of \citet{1998SSRv...85..161G}.  However, our result is in excellent
agreement with that of \citet{2004A&A...415..155B}, who found $\log 
N(\mathrm{O}) = 8.71$ using spectral synthesis along with a 1D LTE MARCS model 
atmosphere and the updated $\log gf$ of \citet{2003ApJ...584L.107J} for 
the isotopic components of the Ni blend.  

More impressive is the agreement with the recent determinations of 
\citet{2001ApJ...556L..63A} and of \citet{2004A&A...417..751A}, both of which 
make use of a three-dimensional, time-dependent hydrodynamical model of the 
solar atmosphere.  The value of Allende Prieto et al., $\log N(\mathrm{O}) = 
8.69$, was derived using the [\ion{O}{1}] $\lambda 6300$ line and careful 
treatment of the Ni blend.  While their $\log gf$ value is 0.20 dex lower 
(translating into a $+0.09$ dex higher O abundance) than the equivalent single 
value of \citet{2003ApJ...584L.107J}, the $\chi^2$-analysis of Allende Prieto 
et al. is sensitive only to the product of $N(\mathrm{Ni}) \times gf$.  Asplund 
et al. obtained highly consistent O abundances from [\ion{O}{1}], permitted 
\ion{O}{1}, OH vibration-rotation, and OH pure rotation lines, finding a mean 
value of $\log N(\mathrm{O}) = 8.66$; their analysis is the first to produce 
consistent solar O abundances derived from the various features used.  The 
Asplund et al. analysis of the [\ion{O}{1}] $\lambda 6300$ feature was 
identical to that of \citet {2001ApJ...556L..63A} but included the Johansson et 
al. (2003) isotopic components of the blending Ni line.  Inclusion of the 
isotopic components altered the theoretical Ni line profile and also the 
$\chi^2$-analysis, although only slightly.  The resulting O abundance as 
derived from the [\ion{O}{1}] $\lambda 6300$ was again $\log N(\mathrm{O}) = 
8.69$.

The concordance of the O abundance found here with those of the 3D analyses of 
\citet{2001ApJ...556L..63A} and \citet{2004A&A...417..751A} is striking given
the use of a 1D analysis of the present study.  This provides sound support 
that the blending Ni feature has been modeled properly and that the {\sf blends}
driver within the {\sf MOOG} package is a robust tool in the analysis of 
blended spectral lines.

Comparing the solar O abundance derived here with that of KH96 it is seen that 
the latter's value is 0.24 dex larger, even though their computed equivalent 
width is $0.1 \; \mathrm{m\AA}$ smaller than ours.  Differences between the two 
studies include the $gf$-values of the [\ion{O}{1}] and blending \ion{Ni}{1} 
lines that were adopted.  KH96 used the sound [\ion{O}{1}] value of 
\citet{1978MNRAS.182..249L}, $\log gf = -9.75$, which does not differ 
significantly from the modern value of $\log gf = -9.717$.  The blending 
$\lambda 6300.34$ Ni line had not been as well studied before the KH96 
analysis, and they assumed $\log gf = -3.00$ based on an inverted abundance 
analysis using the solar Ni abundance ($\log N(\mathrm{Ni}) = 6.25$, the same 
used in the present analysis) of \citet{1989GeCoA..53..197A}.  We are able to 
reproduce the synthesis result of KH96 using their $\lambda 6300$ equivalent 
width and $gf$-values, along with an OVER solar model characterized by their 
adopted atmospheric parameters, with {\sf MOOG} and the {\sf blends} driver.  
Changing the KH96 $\lambda 6300.34$ Ni $gf$-value to the newer isotopic values 
of \citet{2003ApJ...584L.107J} resulted in a derived abundance of 
$\log N(\mathrm{O})=8.71$, almost in perfect agreement with our derived 
abundance of $\log N(\mathrm{O})=8.69$.  For completeness, we point out 
including the modern [\ion{O}{1}] $gf$-value results in $\log 
N(\mathrm{O})=8.68$.  The 0.01 discrepancy between this value and our related 
OVER value is due to the combined small differences in \teff, $\xi$, and EWs 
between the two studies.

\subsubsection{The Dwarfs}
Choosing an accurate C abundance for the dwarfs proved to be problematic due to 
the paucity of existing data and to the lack of measurable C lines in our 
spectra for all the dwarfs save three.  \citet{1978ApJ...223..937T} derived CNO 
abundances of two Hyades mid-F dwarfs, 45 Tau and HD 27561; carbon abundances 
were determined from the group of high-excitation lines in the 7115 {\AA} 
region, as well as two high-excitation lines near 6600 {\AA}.  Tomkin \& 
Lambert found $[\mathrm{C/H}] = +0.07 \mathrm{and} +0.18$ for 45 Tau and HD 
27561, respectively.  \citet{1990ApJ...351..480F} analyzed many of the same 
lines as Tomkin \& Lambert in 13 Hyades F dwarfs, and they find a weighted mean 
of $[\mathrm{C/H}] = +0.04 \pm 0.07$, with individual relative abundances 
ranging from $-0.09 \; \mathrm \; +0.18$.  The two Hyades dwarfs analyzed by 
Tomkin \& Lambert were included in the Friel \& Boesgaard study, for which 
Friel \& Boesgaard find $[\mathrm{C/H}] = -0.09$ for 45 Tau and $[\mathrm{C/H}] 
= +0.03$ for HD 27561.  The study-to-study difference
of $\sim 0.15$ for these two stars is within error estimates for both studies, 
as is the difference in the mean abundances.  \citet{1999A&A...351..247V}
analyzed C in the spectra of 10 Hyades F-type stars and found a mean abundance 
relative to the Sun of $\mathrm{[C/H]} = +0.02$, with individual stellar 
abundances ranging from -0.25 to +0.20 dex and abundance uncertainties as high 
as 0.30-0.40 dex!  We note that Varenne \& Monier also derived C abundances for 
12 Hyades A stars, but given the large uncertainty associated with deriving 
abundances for A-type stars, these are not included in the discussion here.  We 
measured two high-excitation \ion{C}{1} lines (6587.61 and 6655.51; $\chi = 
8.54 \; \mathrm{eV}$) in the spectrum of HIP 19148 ($T_{\mathrm{eff}} = 5978 
\mathrm{K}$) and one of these lines in the spectra of HIP 14976 
($T_{\mathrm{eff}} = 5487 \mathrm{K}$) and HIP 19793 ($T_{\mathrm{eff}} = 5722 
\mathrm{K}$); the resulting mean abundance is $\mathrm{[C/H]} = +0.17 \; 
\pm0.02$, which is in relatively good agreement with the mean result of 
\citet{1978ApJ...223..937T}.  Nonetheless, the modest sample size of Tomkin \& 
Lambert and the sizable spread in the C abundances of 
\citet{1990ApJ...351..480F} and \citet{1999A&A...351..247V} leaves an accurate 
C abundance determination for the Hyades dwarfs wanting.  Furthermore, the 
growing evidence of ionization (\textsection 3.3; Yong et al. 2004) and 
excitation potential-related trends (Schuler et al. 2005; Schuler et al. 2004) 
among cool cluster dwarfs suggests caution when interpreting C 
abundances from the high-excitation \ion{C}{1} lines.  Abundance determinations 
utilizing the $\lambda 8727$ [\ion{C}{1}] feature for Hyades (and other 
clusters) dwarfs are certainly warranted.  In light of the sparse and varying C 
abundance data for the Hyades, a C abundance scaled with the model Fe abundance 
has been chosen here, i.e., $[\mathrm{C/H}] = +0.13$.  Changing this input 
abundance by 0.15 dex results in a moderate change of no more than 0.07 dex in 
the final dwarf O abundances.

Measuring the $\lambda 6300$ [\ion{O}{1}] line strength in the spectra of MS
dwarfs accurately is difficult because of the uncertainty in continuum placement
and the smallness of the feature ($\sim 5.5 \; \mathrm{m\AA}$ for the 
Sun), and measurement uncertainties can be exacerbated if the feature is not 
well-shaped in a spectrum, due to low S/N for instance.  Spectra of the 6300 
{\AA} region of our program stars are shown in Figure 4, where despite the high 
S/N ratio of the spectra, line irregularities can be seen for HIP 14976, HIP 
19148, HIP 19793, HIP 20082, and HIP 22654.  The irregularities are not uniform 
and vary in severity from star-to-star, making it difficult to determine if 
there is systematic cause, e.g., the data reduction process, and if it is 
common to the sample.  In addition to the line shape irregularities, six of the 
spectra obtained with VLT/UVES are doppler shifted such that the $\lambda 6302$ 
telluric line falls on top of the \ion{Sc}{2} line at 6300.68 {\AA}.  
Unfortunately, spectra of telluric standards, i.e., rapidly rotating B or A
stars, were not obtained during the VLT observations, and the telluric lines 
could not be divided out.  However, it appears that the [\ion{O}{1}] line is 
not affected and can be measured cleanly, but the possibility of minor 
contamination cannot be ruled out.  On a more encouraging note, HIP 20082 was 
included in the KH96 sample, allowing us to compare the EW measured from our 
observed spectrum to that measured by KH96 in their synthesized spectrum.  
Happily, our EW measurement of 7.7 m{\AA} is in excellent agreement with the 
KH96 measurement of 7.6 m{\AA}.  Nonetheless, in an effort to account for the 
possible systematic errors related to the irregular line shapes in our spectra, 
an adopted uncertainty of 1 m{\AA} in the EW measurements has been adopted for 
all the dwarfs.  Uncertainties due to individual input parameters affecting the 
[\ion{O}{1}]-based O abundances, as well as the total internal uncertainty, are 
listed in Table 7.  The uncertainties are nearly identical for the OVER and 
MLT5 models, and those listed in Table 7 are from the OVER calculations.  
Uncertainties due to \teff, $\log g$, and $\xi$ are based on the O abundance 
sensitivities to these parameters as discussed in \textsection 3.3.  EW-based 
errors have been calculated by altering the accepted EW measurement by 1.0 
m{\AA}.  
\marginpar{Fig.~4}
\marginpar{Tab.~7}

\subsubsection{The Giants}
There have been no investigations of C abundances in the Hyades giants since 
KH96, who adopted an abundance of $[\mathrm{C/H}] = -0.12$ based on previous 
determinations, and we use this value for the giants here.  A change of 0.15 dex
in the C input abundance results in a modest O abundance difference of 0.03 
dex.  The final O abundances for the giants are listed in Table 6, and the
uncertainties are given in Table 7 and are calculated identically as those for 
the dwarfs, including the adoption of an EW uncertainty of 1 m{\AA}.  The 
abundances are derived using the mean EW calculated from measurements of the 
spectra obtained with the 2.7-m and 2.1-m McDonald telescopes.  The lack of a 
significant difference in the giant O abundances derived using the OVER and 
MLT5 models is consistent with the dwarf results, and we will henceforth 
refer to the OVER results only.

\section{DISCUSSION}
\subsection{O in Hyades Dwarfs}
The individual dwarf O abundances, as well as those for the giants, are plotted 
with error bars in Figure 5.  While the star-to-star dwarf abundances are in 
decent agreement within uncertainties, the plot reveals an apparent increase in
O abundance with decreasing \teff for the three coolest dwarfs of the sample.  
Indeed, the trend is significant at the $\sim 99.5 \%$ confidence level 
according to the linear correlation coefficient.  Such a O abundance trend is
not expected for open cluster stars, which are believed to be chemically
homogeneous.  In order to confirm the reality of the increasing O abundances 
with decreasing \tef, the expected line strengths of the $\lambda 6300$ 
feature, along with the observed EWs from Table 6, are plotted versus \teff in 
Figure 6.  The expected line strengths have been determined by constructing 
synthetic spectra characterized by a single O abundance of $\log N(O) = 8.83$ 
(the mean of the dwarfs with $T_{\mathrm{eff}} > 5000 \; \mathrm{K}$) and with 
Ni abundances of $\mathrm{[Ni/H]} = +0.13$ and $\mathrm{[Ni/H]} = +0.25$, the 
approximate Ni abundances of the six warmest and the three coolest stars, 
respectively (Table 5).  Model atmospheres with \teff in the range $4000 
\leq T_{\mathrm{eff}} \leq 6200 \; \mathrm{K}$, and with $\log g$ and $\xi$ 
calculated as described in \citet{trip}, were interpolated from the OVER grids 
and used for the construction of the synthetic spectra.  EWs of the synthesized 
$\lambda 6300$ features were measured using {\sf SPECTRE} in the same manner as
the observed lines.  Excellent to good agreement exists between the observed 
and synthesized (with $\mathrm{[Ni/H]} = +0.13$) EWs for the six warmest stars 
of the sample.  The observed EWs of the three coolest dwarfs start to deviate 
from the synthesized measurements, with the observed EW of the coolest Hyad, 
HIP 22654, $\sim 2.2 \; \mathrm{m\AA}$ larger than expected (with 
$\mathrm{[Ni/H]} = +0.25$); this deviation for HIP 22654 is 2.2 times the 
adopted line strength uncertainty.  These data support the notion that the O 
abundance increase among the cool dwarfs is statistically significant.  

A growing body of evidence is emerging that suggests spectroscopic abundance 
analyses of cool ($T_{\mathrm{eff}} \lesssim 5500 \; \mathrm{K}$) MS dwarfs 
making use of 1D, plane-parallel model atmospheres are unable to determine 
accurately the photospheric abundances of certain elements for these stars, 
with the abundance anomalies apparently increasing with decreasing \tef.  
Examples include the \ion{Fe}{2}-\ion{Fe}{1} abundance discrepancies in the 
Hyades (\textsection 3.3; Yong et al. 2004), M34 \citep{2003AJ....125.2085S}, 
and UMa \citep{ksuma}; the high-excitation \ion{O}{1} triplet-based O abundance 
trends in the Hyades \citep{trip}, the Pleiades and M34 
\citep{2004ApJ...602L.117S}, UMa \citep{ksuma}, and chromospherically active 
stars \citep{2004A&A...423..677M}; and excitation-related abundance differences 
between two K-dwarfs and the rest of the sample of nine M34 dwarfs 
\citep{2003AJ....125.2085S}.  The results of all of these studies point to the 
effects of overionization and excitation, possible phenomena in cool dwarfs 
that were first recognized by \citet{1998A&AS..129..237F}.  However, the 
[\ion{O}{1}] $\lambda 6300$ feature results from a ground-state, magnetic 
dipole transition with a very weak electric quadrupole contribution, and its 
strength should be impervious to overionization/excitation effects.  What, 
then, could be the cause of the apparent \tef-dependent increase in 
[\ion{O}{1}]-based O abundances among the cool dwarfs of our sample?
\marginpar{Fig.~5}
\marginpar{Fig.~6}

In an attempt to answer this question, attention is first turned to the 
blending Ni line(s) at 6300.34 {\AA}, which results from a transition with a
moderately-high, lower excitation potential of 4.27 eV and may be affected by 
the overexcitation effects discussed above.  Care was taken to derive an 
accurate Ni abundance for each star, so that the contribution of the Ni blend 
to the measured EW of the $\lambda 6300$ line could be determined.  The Ni
lines chosen for measurement have lower excitation potentials in the range 
$1.83 \leq \chi \leq 4.42 \; \mathrm{eV}$, with all but one line having 
$\chi \geq 3.54 \; \mathrm{eV}$.  The absolute Ni abundances of the warmest 
star in the sample (HIP 19148; 5978 K) and of the coolest star in the sample 
(HIP 22654; 4573 K) are plotted as a function of $\chi$ in Figure 7a.  The Ni 
abundances of HIP 19148 are independent of $\chi$, while those for HIP 22654 
derived from high-excitation lines are $\sim 0.15 \; \mathrm{dex}$ higher than 
that derived from the low-excitation line, evincing clearly $\chi$-dependent 
abundances in at least the coolest star.  In response, the relative Ni 
abundance of each star has been re-derived from a set of 3-5 Ni lines with 
$\chi$ ranging from 4.23-4.26 eV (Table 8); the Ni abundance from 
these lines will presumably provide a better indication of the line strength of 
the 4.27 eV Ni line blend and thus of its contribution to the $\lambda 6300$ 
EW.  The resulting mean abundances, along with the originally accepted [Ni/H] 
abundances, are plotted for each star versus \teff in Figure 7b.  Although an 
unequivocal \tef-dependent increase in Ni abundances with decreasing \teff is 
apparent, no appreciable difference between the two abundance measures is seen. 
Thus, the [\ion{O}{1}] abundance trend does not appear to be the result of the 
Ni blend and overexcitation effects.
\marginpar{Fig.~7}
\marginpar{Tab.~8}

Another possible cause of the increase in [\ion{O}{1}] abundances in the cool
dwarfs could be related to overionization/excitation effects, namely an actual
increase in the number of O atoms in the photospheres of the cool dwarfs due 
to greater than predicted dissociation of O-containing molecules, especially 
CO, the most prominent O-containing molecule in the photospheres of our program
stars.  The larger than predicted EWs of the $\lambda 6300$ [\ion{O}{1}] line
measured in the spectra of the cool dwarfs would be a natural consequence of an 
increased number of O atoms that would result from the dissociation of 
molecules, and not accounting for this in the abundance derivations would lead
to larger than expected abundances.  The plausibility of this scenario has been 
investigated by rederiving the O abundances of each dwarf and excluding CO from 
the molecular equilibrium calculations performed by {\sf MOOG}, effectively 
simulating the complete dissociation of CO in the stellar photospheres.  
Astonishingly, the resulting O abundances of the cool stars are brought into 
excellent agreement with those of the warm stars, as can be seen in Figure 8.  
The mean of these newly derived abundances is $\mathrm{[O/H]} = 0.12$ with a 
standard deviation of only 0.03 dex.   The results of this exercise in 
themselves are of little scientific value; however, they do provide significant 
motivation to investigate O abundances derived from molecular spectral lines: 
if \tef-dependent overdissociation of O-containing molecules is occurring in 
the photospheres of cool dwarfs, one would expect to see O abundances derived 
from molecular lines exhibit a trend of decreasing abundances with decreasing 
\tef, in contrast to the increasing abundances with decreasing \teff observed 
in the O abundances derived from atomic lines.
\marginpar{Fig.~8}

Accordingly, O abundances of four Hyades dwarfs (HIP 19148, HIP 19793, HIP
21099, and HD 29159) have been derived via spectral synthesis of the near-UV, 
$\lambda 3167$ OH line ($\chi = 1.11 \; \mathrm{eV}$) in high-quality 
($\mathrm{S/N} \approx 100$, $R \approx 45,000$) spectra obtained with the Keck 
I telescope and HIRES spectrograph.  Abundances derived from molecular species 
are highly temperature sensitive, and  molecular lines in the UV may be 
affected by NLTE effects \citep{1975MNRAS.170..447H} and contributions of
metals to the continuous opacity \citep{2003ApJ...591.1192A}.  Nonetheless,
LTE analyses of near-UV OH lines have been shown to produce reliable results 
\citep{1998ApJ...507..805I}.  For the present study, using OH ought to be a 
good choice to test for the effects of overdissociation because (1) it has a 
lower dissociation energy ($D_0 = 4.39 \; \mathrm{eV}$) compared to that of the 
more multitudinous CO ($D_0 = 11.09 \; \mathrm{eV}$); and (2) even a small 
amount of overdissociation will affect significantly the observed line 
strengths due to the relatively lower number abundance of OH molecules in the 
atmospheres of late-G and early-K dwarfs.  The near-UV spectra of HIP 19148, 
HIP 19793, and HIP 21099 are those from the study of \citet{2002ApJ...565..587B}, and the data 
are fully described therein.  The near-UV spectrum of HD 29159 was obtained 
on 2004 Dec 15 with Keck I and HIRES, using the ``b'' configuration; a 
resolution of $R \approx 45,000$ and a $\mathrm{S/N} \approx 80$ at 3167 {\AA} 
were achieved.  The observations made use of the new three-chip ($2048 \times 
4096$ each) CCD mosaic.  Identifying clean spectral lines in the near-UV is 
difficult due to the considerable number of identified and unidentified 
blends.  The selection of OH lines compiled by the related studies of 
\citet{1998ApJ...507..805I} and \citet{2001ApJ...551..833I} was used as a 
starting point for identifying acceptable lines in our spectra, and the 
$\lambda = 3167.169 \; \mathrm{\AA}$ line was found to be the most promising 
candidate.  While this line is surely contaminated by blends, OH is the 
dominant contributor, and the feature should provide a good indication of
{\it relative} abundance differences, if present.  It should be stressed that 
we are not attempting to determine accurate O abundances from the analysis of 
the OH feature; rather, the goal is to determine if {\it relative}, star-to-star
abundance differences are observed.  The spectral synthesis was carried out with
{\sf MOOG} and an initial line list compiled from the VALD database.  Final
$gf$-values for the OH and surrounding lines were determined by fitting the
region for the Sun using the Kurucz solar atlas flux spectrum 
\citep{1984sfat.book.....K} and adopting an input O abundance of $\log N(O) = 
8.69$, the solar abundance derived from the $\lambda 6300$ [\ion{O}{1}] line in 
our high-quality McDonald 2.7-m spectrum of the daytime sky.  The Kurucz solar
spectrum was smoothed by convolving it with a Gaussian profile with a FWHM of 
0.052 {\AA} \citep{1995A&A...302..184G}.

Synthetic fits to the observed spectra of the four Hyads are quite 
satisfactory and are presented in Figure 9.  The synthetic spectra have been
smoothed using Gaussians with FWHM values determined via a cross-correlation
analysis using the {\sf fxcor} utility within {\sf IRAF}.  Numerous synthetic
spectra of the 3167 {\AA} region for a fiducial Hyades dwarf having $T_{\rm eff}
= 5500 \; {\rm K}$ and ${\rm [Fe/H]} = +0.13$, each broadened with a Gaussian of
differing FWHM, were cross-correlated with the observed Kurucz solar spectrum 
as a template.  The resulting cross-correlation function (CCF) widths were then
plotted versus the input FWHM smoothing values of the synthetic Hyades spectrum,
and a fourth order polynomial was fit to this relation.  Our observed Keck/HIRES
Hyades spectra were then cross-correlated with the Kurucz solar spectrum, the 
CCF width measured, and the appropriate FWHM determined from our synthetic 
relation.  In Figure 9, the final absolute O abundance that resulted in the 
best fit for each star is included, as well as syntheses for $\log N(O) = 8.46$ 
and $\log N(O) = 9.36$.  Close inspection reveals a gradual migration of the 
observed OH line from near the synthesis characterized by the low O abundance 
($\log N(O) = 8.46$) in Figure 9a (HIP 19148, the warmest star) toward the 
synthesis characterized by the high O abundance ($\log N(O) = 9.36$) in Figure 
9d (HD 29159, the coolest star), suggesting an increase in the O abundance with 
decreasing \tef.  This behavior is confirmed in Figure 10, where the OH-derived 
O abundances are plotted versus \tef, and is contrary to the expected result if 
overdissociation of O-containing molecules is occurring.  This leads to the 
conclusion that the increasing [\ion{O}{1}] abundances among the cool dwarfs is 
not the result of overdissociation of O-containing molecules, although 
additional OH data for Hyades dwarfs cooler than $\sim 5000 \; \mathrm{K}$ are 
desirable for greater confirmation or alternatively, to determine if the 
abundances continue to increase at lower \tef.
\marginpar{Fig.~9}
\marginpar{Fig.~10}

The nature of the observed increase in O abundances as derived from the $\lambda
6300$ [\ion{O}{1}] feature remains uncertain.  Further study of this phenomenon 
will benefit greatly from additional data for dwarfs with \teff as cool as 4200 
K, where the expected EWs of the [\ion{O}{1}] feature for Hyades metallicity 
stars are $> 3 \; \mathrm{m\AA}$ (Figure 4) and should be measurable in 
high-quality spectra.  Consequently, if the EWs are indeed enhanced in the 
spectra of dwarfs cooler than those study here, then future analyses could be 
extended to even cooler \tef.  Regardless, a caveat must be extended: current 
abundance derivations for cool dwarfs, irrespective of the lower excitation 
potential of the spectral lines being employed, should be received with caution.
Hyades dwarf O abundances as derived from the $\lambda 6300$ [\ion{O}{1}], 
near-IR $\lambda 7774$ triplet, and the $\lambda 3167$ OH line are provided in 
Figure 10.  The [\ion{O}{1}] and OH-based dwarf abundances of the warmest stars 
agree nicely.  The triplet-based dwarf O abundances of the warm stars are 
offset by $\sim 0.10$ dex (Figure 5), similar to the $\sim0.10$ dex difference 
in the \ion{Fe}{2}-\ion{Fe}{1} abundances in the warmest stars (Figure 3).  The 
reason for the offsets is not clear but might be related to NLTE effects or 
undetermined internal or systematic errors \citep{trip}.  Below about 5000 K, 
the [\ion{O}{1}] and triplet abundance trends clearly diverge, but both show 
signs of increasing with decreasing \tef.  Previous, as well as the current, 
studies have reported abundance trends among cool cluster dwarfs that present a 
coherent picture of overionization/excitation related effects (Figure 3, Figure 
7, Schuler et al. 2005, Yong et al. 2004, etc.), and \citet{trip} 
have demonstrated the plausibility that the effects might be the result of
temperature inhomogeneities due to the presence of photospheric spots, faculae, 
and/or plages.  The [\ion{O}{1}]-derived abundance trend presented here does 
not fit into this scenario, and if confirmed, additional mechanisms to explain 
the behavior will be required.  

Conversely, the O abundances of the six warm dwarfs ($5075 \leq 
T_{\mathrm{eff}} \leq 5978$) in our sample are in excellent agreement and do 
not engender incertitude in their accuracy.  Using the abundances of these six 
warm dwarfs, excluding those of the three coolest due to their uncertainty, an 
O abundance of $[\mathrm{O/H}] = +0.14 \; \pm 0.02$ (uncertainty in the mean) 
is found for the Hyades cluster.  This abundance agrees near-perfectly with 
that found by KH96, $[\mathrm{O/H}] = +0.15 \; \pm 0.01$; although the 
agreement is difficult to explain.  HIP 20082 ($T_{\mathrm{eff}} = 4784$), the 
one star in common to KH96 and the present study, is not included in the 
cluster mean here.  An absolute abundance of $\log N(O) = 9.09$ for HIP 20082 
was derived by KH96, whereas we find $\log N(O) = 8.96$; however, the absolute 
solar abundances of the two studies differ by 0.24 dex, resulting in a 
study-to-study difference of 0.11 dex in the relative O abundances for HIP 
20082.  This difference can be attributed to the combined effects of small 
differences in stellar parameters, EWs, and sensitivities to changes in the 
$\log gf$-values of the [\ion{O}{1}] and Ni lines.  Thus, it seems the analysis 
of each study has conspired to provide the same cluster abundance despite the 
difference in the relative abundances for the one star in common.

\subsection{O in Hyades Giants}
The O abundances derived for the Hyades giants are in good star-to-star
agreement (Table 6) and have a mean value of $\mathrm{[O/H]} = +0.08 \; \pm 
0.02$ (uncertainty in the mean).  This is 0.16 dex higher than the mean giant 
abundance obtained by KH96; the difference can be attributed to the complex 
interplay of the discordance in adopted solar abundances, the updated atomic 
parameters for the [\ion{O}{1}] and Ni blend lines, and small disagreements in 
EWs.  The differences between our measured EWs and those used by KH96, which 
are means of the values of \citet{1981ApJ...248..228L}, 
\citet{1982A&A...115..145K}, and \citet{1985A&A...148..105G}, average about 2.9 
m{\AA} and are within the combined uncertainties.

Oxygen abundances derived from the near-IR, high-excitation $\lambda 7774$
\ion{O}{1} triplet for both unevolved and evolved stars alike are known to be
influenced by NLTE effects (e.g., King \& Boesgaard 1995; Cavallo, Pilachowski,
\& Rebolo 1997).  Nevertheless, deriving O abundances from these lines can be
informative.  EWs of the triplet as measured in the high-quality McDonald 2.7-m 
spectra of the three Hyades giants are given in Table 9.  Contrary to 
expectations, the EW of the central line of the triplet ($\lambda = 7774.16 \;
\mathrm{\AA}$) is larger than the blue line ($\lambda = 7771.95 \; 
\mathrm{\AA}$), suggesting the presence of a blend that is otherwise negligible 
for dwarfs.  No asymmetry of the central line is seen in the McDonald 
spectra (Figure 11), but a blending feature might be present, possibly a 
\ion{Fe}{1} line at 7774.00 {\AA} \citep{1998PASJ...50...97T}.  LTE abundances 
(Table 9) were derived utilizing the same model atmospheres as for the 
[\ion{O}{1}] analysis; atomic parameters for the triplet are those used by 
\citet{trip}.  Excluding the central line from further analysis, the giant LTE 
mean O abundances derived from the remaining two lines of the triplet are 
approximately 0.28 dex higher than those derived from the [\ion{O}{1}] line,
qualitatively in-line with NLTE calculations.  NLTE corrections to the LTE
abundances have been estimated using the results of \citet{2003A&A...402..343T}
who performed extensive NLTE calculations for a large collection of model
atmosphere parameters for late-F through early-K stars.  The NLTE corrections
are available for only discrete steps in \tef, $\log g$, and $\xi$, and
thus a mean of the corrections from two separate grid steps has been taken 
here.  The corrections for the Hyades giants have been determined using the 
values for the parameter sets of $T_{\mathrm{eff}} = 5000 \; \mathrm{K}$, 
$\log g = 2.00$, $\xi = 1.00 \; \mathrm{km} \; \mathrm{s}^{-1}$ and 
$T_{\mathrm{eff}} = 5000 \; \mathrm{K}$, $\log g = 3.00$, $\xi = 1.00 \; 
\mathrm{km} \; \mathrm{s}^{-1}$.  The resulting NLTE abundances are included in 
Table 9, and both LTE and NLTE abundances are plotted in Figure 10.  We 
calculate a mean NLTE abundance, excluding the abundance derived from the 
central line, for the giants of $\mathrm{[O/H]} = 0.17 \; \pm 0.02$ 
(uncertainty in the mean).  This result is in satisfactory accordance with the 
[\ion{O}{1}] giant abundances and adds muster to the super-solar O abundances 
found here. 
\marginpar{Tab.~9}
\marginpar{Fig.~11}

\subsection{Is There a Hyades Giant-Dwarf O Abundance Discrepancy?}
With mean O abundances for both the dwarfs and giants comfortably in hand, we 
can now proceed to address the question of the so-called Hyades giant-dwarf 
oxygen discrepancy initially reported by KH96.  KH96 derived relative 
abundances of 0.15 and -0.08 dex for the dwarfs and giants, respectively, 
resulting in a 0.23 dex O abundance difference.  Our analysis does not confirm 
this large abundance discordance; the mean O abundances derived for the dwarfs 
and giants are 0.14 and 0.08 dex, reducing the difference to only 0.06 dex.  
Furthermore, the abundances of the warm dwarfs ($T_{\mathrm{eff}} > 5000 \; 
\mathrm{K}$) and of the giants are in good star-to-star agreement (Figure 5), 
and the samples are statistically indistinguishable.  We note that the low O
abundance ($\mathrm{[O/H]} = -0.08 \pm 0.01$) derived for the giants by KH96 is 
in near-perfect agreement with the value ($\mathrm{[O/H]} = -0.06 \pm 0.15$) 
obtained by \citet{1993ApJ...412..173G} via an NLTE analysis of the triplet in 
the spectra of 25 Hyades F dwarfs.  While these results present a resolution to 
the giant-dwarf O discrepancy as well, they suggest a lower cluster O 
abundance than found here and by others (King 1993; Schuler et al. 2005; 
Boesgaard 2005).  Given our consistently super-solar O abundances derived from
the various features considered, we believe our values to be more accurate.

The agreement between the dwarf and giant O abundances corroborates the 
conclusion reached from our evolutionary model of a 2.5 $M_{\sun}$ star
described in \S 1.1, i.e., the Hyades giants have not mixed O-depleted material 
into their atmospheres, a possible cause of the giant-dwarf O abundance 
discrepancy suggested by KH96.  Instead, the culprit seems to have been 
systematic errors related primarily to the treatment of the Ni blend of the 
$\lambda 6300$ feature, as described in Sections 3.4.1, 4.1, and 4.2.  This 
reanalysis of the giant-dwarf oxygen discrepancy seemingly demonstrates that 
even careful relative abundance studies can lead to erroneous results.

\section{SUMMARY}
We have analyzed the $\lambda 6300$ [\ion{O}{1}] spectral feature in
high-quality spectra obtained with the VLT Kueyen (UT2) telescope at the 
European Southern Observatory, Paranal and the Harlan J. Smith and Otto Struve 
telescopes at The McDonald Observatory in order to revisit the Hyades 
Giant-Dwarf Oxygen Discrepancy initially reported by 
\citet{1996AJ....112.2650K}.  Abundances of Fe, Ni, and O have been derived for 
nine Hyades dwarfs and three Hyades giants using two sets of model atmosphere 
grids based on the Kurucz ATLAS9 code.  All results are found to be independent 
of model atmosphere, as previously demonstrated for MS dwarfs 
\citep{2004ApJ...602L.117S} but shown for the first time for red giant stars 
here.  Dwarf Fe abundances derived from singly ionized lines are generally 
found to be greater than those derived from neutral lines, with the discrepancy 
increasing with decreasing \tef; this result confirms the original espial by 
\citet{2004ApJ...603..697Y}.  No such discordance is observed for the giants, 
for which ionization balance is achieved within uncertainties.  Mean
Fe abundances from \ion{Fe}{1} lines alone are 0.08 and 0.16 for the dwarfs and
giants, respectively.  Both of these values are in good agreement with previous
determinations for the Hyades.

Our O abundance analysis has improved on that of King \& Hiltgen most 
significantly in the treatment of the \ion{Ni}{1} line that is blended with the
[\ion{O}{1}] feature.  \citet{2003ApJ...584L.107J} has shown experimentally 
that the Ni blend is actually due to two isotopic components, and the weighted
$gf$-value for each component has been subsequently calculated by Bensby et al.
(2004).  The weighted isotopic values have been used here.  Our resulting solar
abundance, based on an equivalent width of the [\ion{O}{1}] line measured in a 
high-quality spectrum of the day-time sky at The McDonald Observatory, is $\log 
N(O) = 8.69$, in excellent agreement with the results of Allende Prieto et al.
(2001) and of Asplund et al. (2004), both of which make use of a
three-dimensional, time-dependent hydrodynamical model of the solar atmosphere. 
The solar abundance of King \& Hiltgen is 0.24 dex greater than that found here;
it is shown that this large difference can be almost entirely attributed to the
updated $gf$-value of the Ni blend.

The dwarf O abundances based on the [\ion{O}{1}] feature are in agreement 
within uncertainties, but an apparent trend of increasing abundances with 
decreasing \teff is observed among the three coolest dwarfs of the sample.  
Possible explanations for the anomalous abundances include a greater than 
expected contribution to the EW of the $\lambda 6300$ feature from the Ni blend 
($\chi = 4.27 \; \mathrm{eV}$) due to overexcitation-related effects and/or an 
increase in the amount of atomic O due to overdissociation of O-containing 
molecules.  Both of these possibilities are investigated, but neither are found 
to be viable.  Analysis of the [\ion{O}{1}] line in the spectra of additional 
cool open cluster dwarfs is recommended to confirm the abundance trend found 
here.  Using the abundances of the six warmest dwarfs in the sample, we find a 
mean value of $\mathrm{[O/H]} = +0.14 \; \pm 0.02$.  Our dwarf O abundance is 
in excellent agreement with that of King \& Hiltgen, despite considerable 
differences in the analyses.

Analysis of the [\ion{O}{1}] feature in spectra of the Hyades giants results in
a mean abundance of $\mathrm{[O/H]} = +0.08 \; \pm 0.02$.  This is 0.16 dex
higher than the value obtained by King \& Hiltgen; the difference is again
mainly due to the updated $gf$-value of the Ni blend.  We also derive O
abundances for the giants from the near-IR, high-excitation triplet.  The NLTE
corrections of \citet{2003A&A...402..343T} have been adopted, and the resulting
mean NLTE abundance is $\mathrm{[O/H]} = +0.17$ and is in satisfactory 
accordance with the giant and dwarf [\ion{O}{1}]-based measurements.  Comparing 
the dwarf and giant mean O abundances, we do not confirm the Giant-Dwarf Oxygen
Discrepancy initially reported by King \& Hiltgen.  The abundance difference
found by King \& Hiltgen seems to have resulted primarily from systematic 
errors related to the $gf$-value of the Ni blend, which was poorly constrained
at that time.

In closing, a summation of the Hyades cluster O abundance derived from the
myriad indicators is in order.  The totality of our results are given in Figure
10, where the dwarf LTE O abundances as derived from the high-excitation 
\ion{O}{1} triplet from \citet{trip} are also plotted.  The 
figure presents a convoluted and possibly discouraging picture with regards to 
O abundance derivations, but all hope is not lost.  The $\lambda 6300$ 
[\ion{O}{1}] line appears to remain the best indicator of O abundances in the 
atmospheres of both dwarfs (black circles) and giants (black triangles), 
although some monition is required for dwarfs with $T_{\mathrm{eff}} \lesssim 
5000 \; \mathrm{K}$.  Abundances derived from the high-excitation triplet 
exhibit rich behavior in both dwarfs (red circles) and giants (red triangles).  
For the dwarfs with $T_{\mathrm{eff}} \gtrsim 6100 \; \mathrm{K}$ and the 
giants, the LTE abundances are in accord with current NLTE expectations 
\citep{2003A&A...402..343T}.  Indeed, applying the NLTE corrections 
appropriate for the Hyades giants from the extensive collection of 
\citet{2003A&A...402..343T} brings their triplet abundances into close 
agreement with their [\ion{O}{1}] abundances.  The behavior of the LTE triplet 
abundances of the cool dwarfs ($T_{\mathrm{eff}} \lesssim 5450 \; \mathrm{K}$), 
on the other hand, are not predicted by current NLTE calculations; \citet{trip}
showed that temperature inhomogeneities due to spots, faculae, and/or plages
might be a plausible explanation for the cool dwarf O abundance anomalies.  At 
intermediate \teff ($5450 \lesssim T_{\mathrm{eff}} \lesssim 6100 \; 
\mathrm{K}$), no discordance between triplet and [\ion{O}{1}] abundances is 
expected, especially if the abundances are given relative to solar.  As 
discussed in \textsection 4.1, the significance of the
$\sim 0.10$ dex offset between the abundances from these two indicators is not
clear, but may be related to NLTE effects or systematic errors.  Relatedly,
Boesgaard (2005) recently derived O abundances using an LTE analysis of the
triplet for 18 Hyades dwarfs in the intermediate \teff range given above and
found a mean abundance of $\mathrm{[O/H]} = +0.17 \pm 0.01$, a value
intermediate to the triplet abundance of \citep{trip} and the [\ion{O}{1}]
abundance found here.  Deriving abundances from a single near-UV OH line (blue
circles) can be precarious given uncertainties associated with ill-constrained 
atomic parameters, high sensitivities to stellar parameters, and continuum 
placement; nevertheless, the trend of increasing OH-based abundances with 
decreasing \teff we find is reminiscent of those based on the triplet and 
[\ion{O}{1}] line.  Finally, combining the [\ion{O}{1}] results for both the 
warm dwarfs ($T_{\mathrm{eff}} \gtrsim 5000 \; \mathrm{K}$) and the giants, a 
cluster abundance of $\mathrm{[O/H]} = +0.12 \; \pm 0.02$ is achieved, giving 
the best indication of the Hyades O abundance.

\acknowledgements
S.C.S. thanks the South Carolina Space Grant Consortium for providing support
through the Graduate Student Research Fellowship.  S.C.S. and JRK gratefully 
acknowledge support for this work by grant AST 02-39518 to J.R.K. from the 
National Science Foundation, NASA support via grant \#HF-1046.01-93A to J.R.K. 
from the Space Science Telescope Institute, as well as a generous grant from
the Charles Curry Foundation to Clemson University.  We are grateful to Prof. 
Ann Boesgaard for obtaining and providing the Keck/HIRES, near-UV spectra of HD 
29159.  We also thank Ms. Abigail Daane for her assistance with the 
observations at the 2.7-m McDonald telescope.  This research has made use of 
the SIMBAD database, operated at CDS, Strasbourg, France.

\newpage

\begin{figure}
\includegraphics[angle=90]{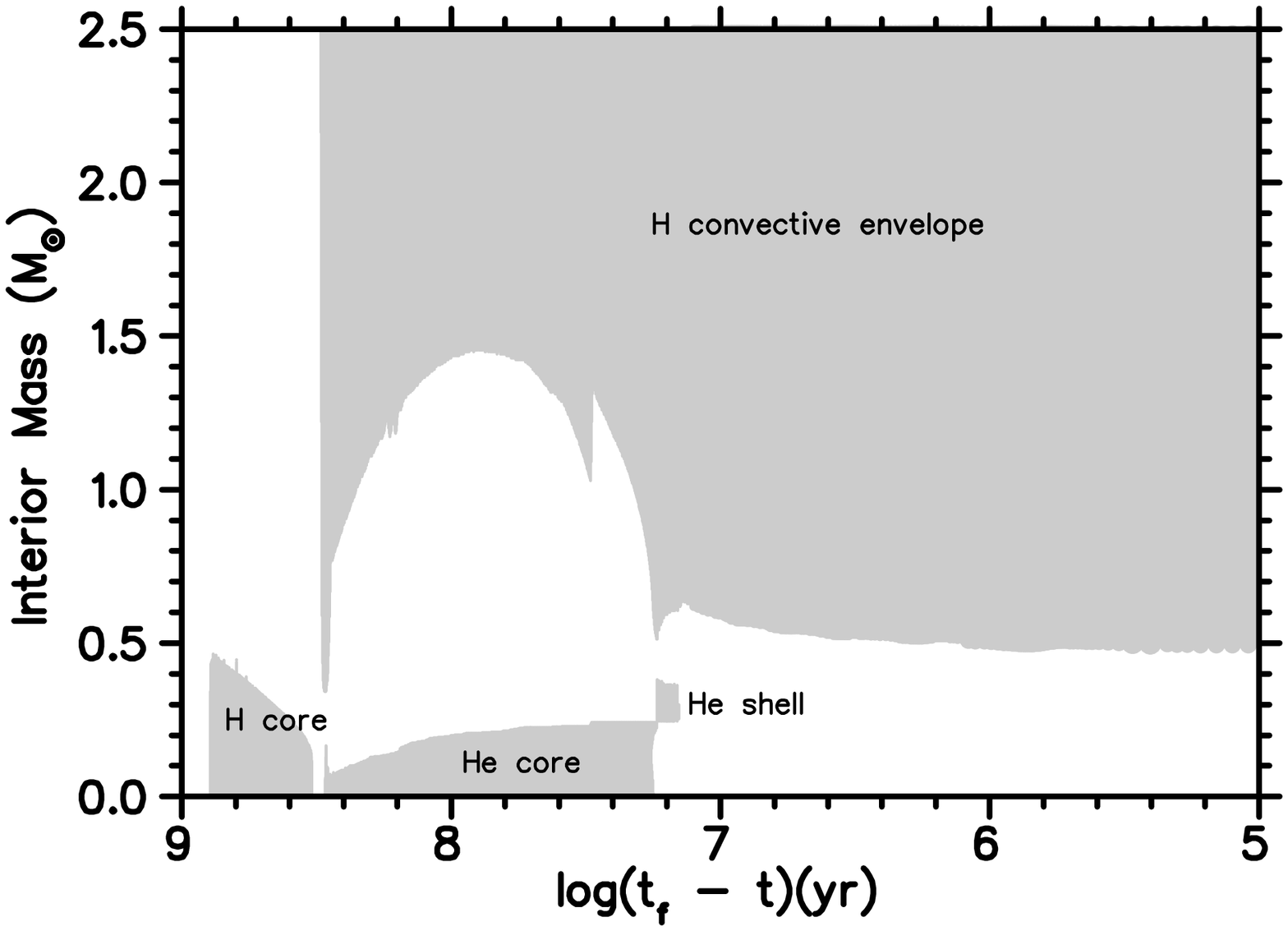}
\caption{Stellar structure as a function of time from the 2.5 $M_{\sun}$ model
with an initial Hyades-like composition.  Time measured as the difference 
between total elapsed time of the model ($\mathrm{t}_{\mathrm{f}} \approx 7.9 
\times 10^8 \; \mathrm{yr}$) and time (t) from $\mathrm{t} = 0$ is plotted on a 
logarithmic scale along the abscissa.  Convective regions are given in gray, 
and radiative regions are white.}
\end{figure}

\begin{figure}
\plotone{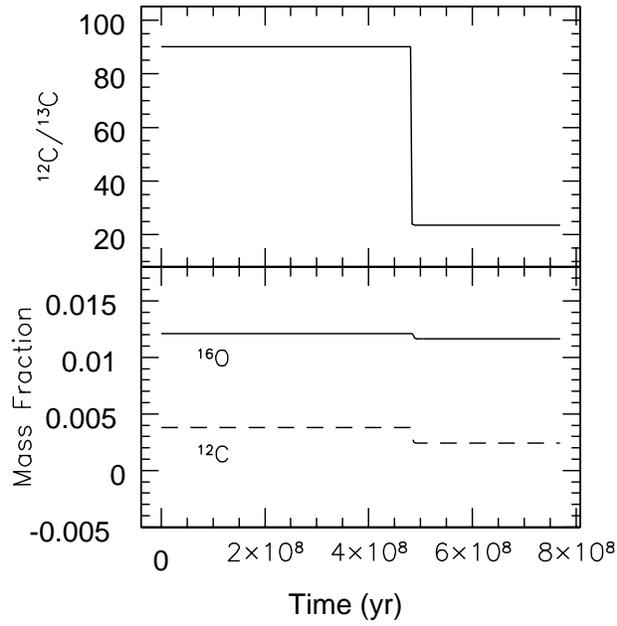}
\caption{Isotopic abundances of the surface of the 2.5 $M_{\sun}$ model with an
initial Hyades-like composition as a function of time.  The top panel shows the
evolution of the \cara/\carbs ratio, while the bottom panel displays the
evolution of the \oxys (solid line) and \caras abundances (dashed line).}
\end{figure}

\begin{figure}
\plotone{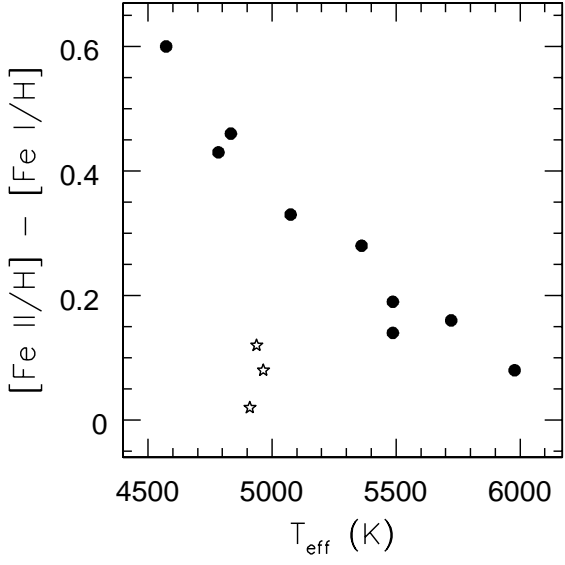}
\caption{Difference in Fe abundances derived from \ion{Fe}{2} and \ion{Fe}{1}
lines plotted against \tef.  The abundance differences for the dwarfs are given
as circles and those for the giants as stars.}
\end{figure}

\begin{figure}
\plotone{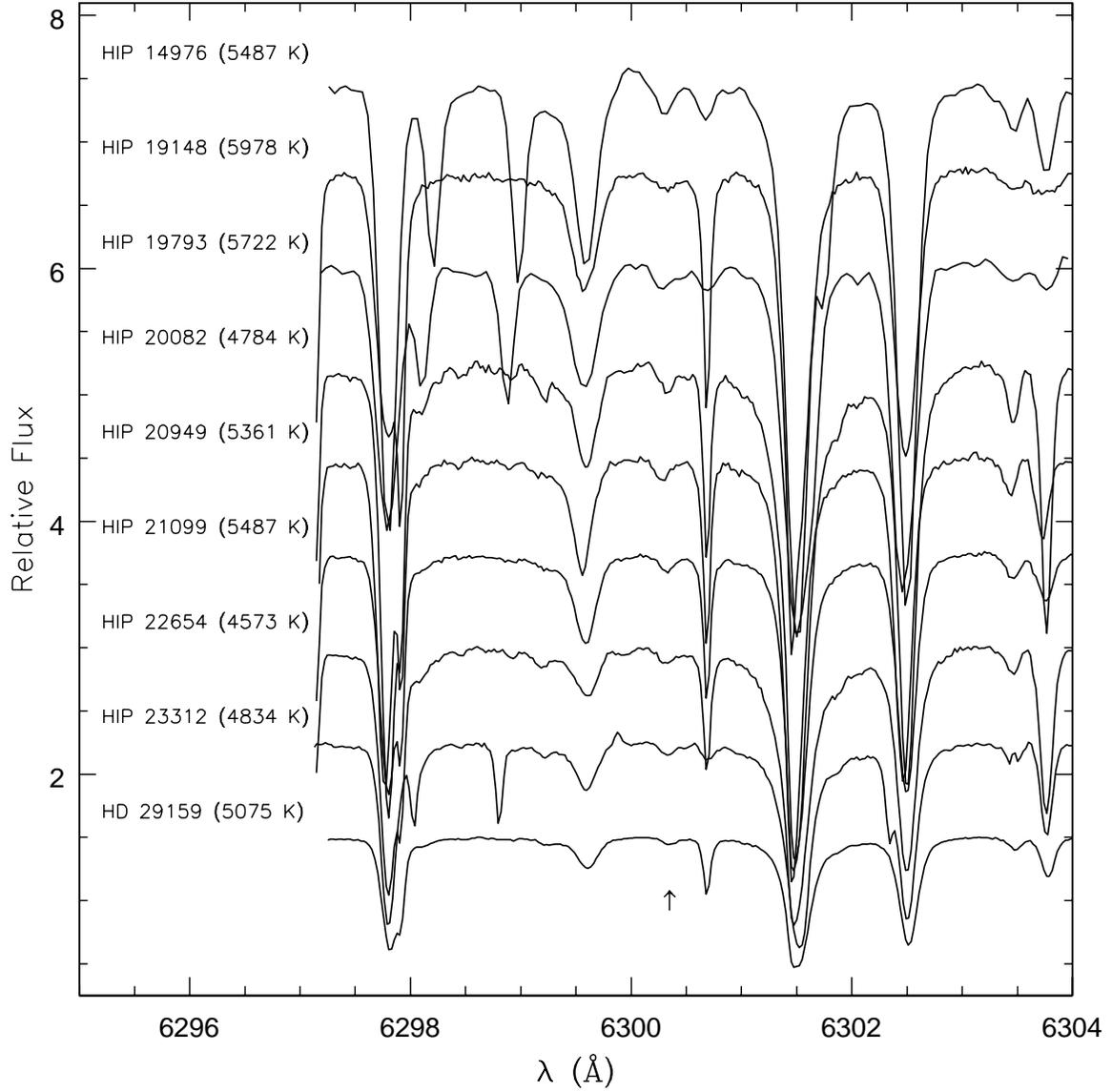}
\caption{VLT/UVES and McDonald 2.7-m/2dcoud{\'e} spectra of the $\lambda 6300$
[\ion{O}{1}] region for the Hyades dwarf sample.  The [\ion{O}{1}] line is
marked by the arrow.}
\end{figure}

\begin{figure}
\plotone{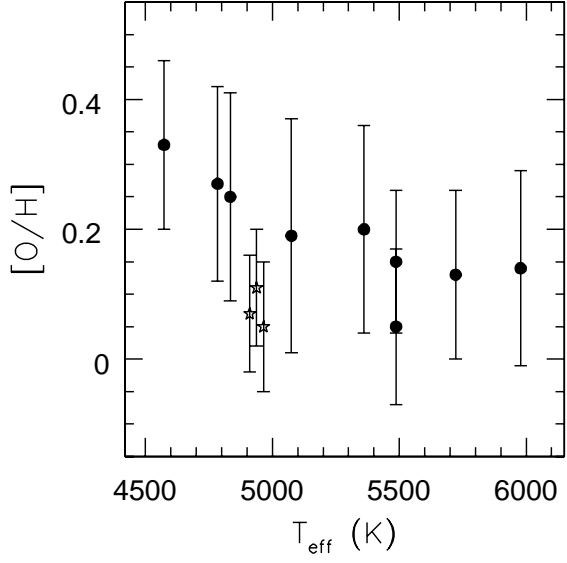}
\caption{Relative LTE O abundances of Hyades dwarfs (circles) and giants 
(stars) versus \tef.  The error bars represent the total internal uncertainties 
as described in the text.}
\end{figure}

\begin{figure}
\plotone{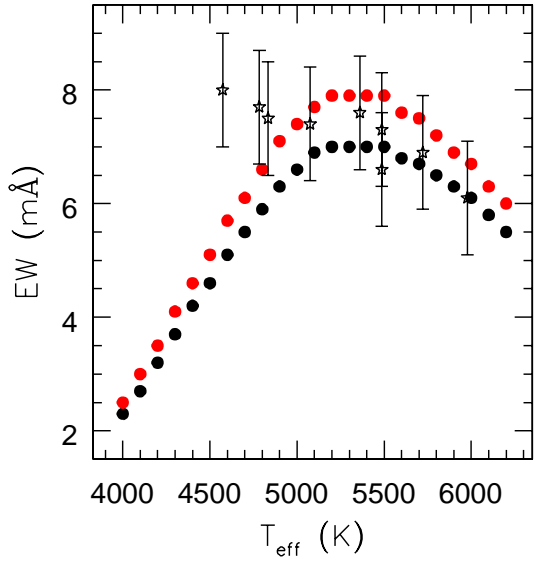}
\caption{Equivalent widths of the $\lambda 6300$ [\ion{O}{1}] feature for the
Hyades dwarf sample versus \tef.  Open stars with error bars are the observed 
EWs.  Closed circles are EWs measured from synthetic spectra with input Ni 
abundances of $\mathrm{[Ni/H]} = +0.13$ (black) and $\mathrm{[Ni/H]}=+0.25$ 
(red).}
\end{figure}

\begin{figure}
\plotone{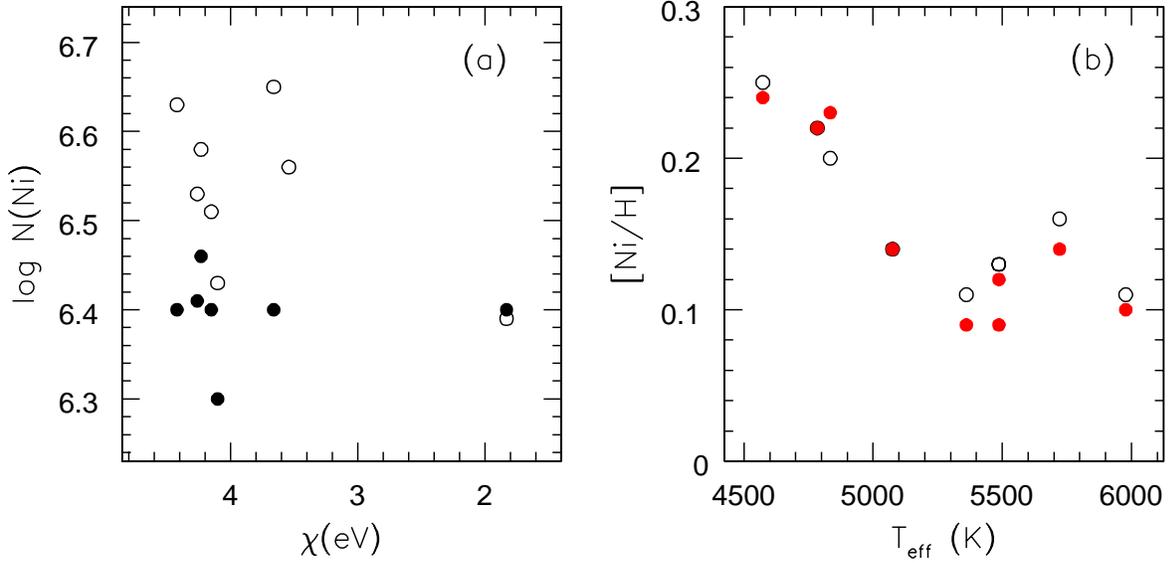}
\caption{(a) Line-by-line absolute Ni abundances plotted against \tef.  Filled 
circles are the abundances for HIP 19148, the warmest dwarf in the sample
($T_{\mathrm{eff}} = 5978 \; \mathrm{K}$); open circles are the abundances of
HIP 22654, the coolest dwarf in the sample ($T_{\mathrm{eff}} = 4573 \; 
\mathrm{K}$).  (b) Relative Ni abundances of the Hyades dwarf sample versus 
\tef.  Ni abundances derived with the original set of lines (Table 2) are given
as open circles; red circles are the abundances derived with a set of Ni lines 
with $4.23 \leq \chi \leq 4.26 \; \mathrm{eV}$, values that are approximately 
equal to that ($\chi = 4.27 \; \mathrm{eV}$) of the \ion{Ni}{1} line blended 
with the $\lambda 6300$ [\ion{O}{1}] feature.}
\end{figure}

\begin{figure}
\plotone{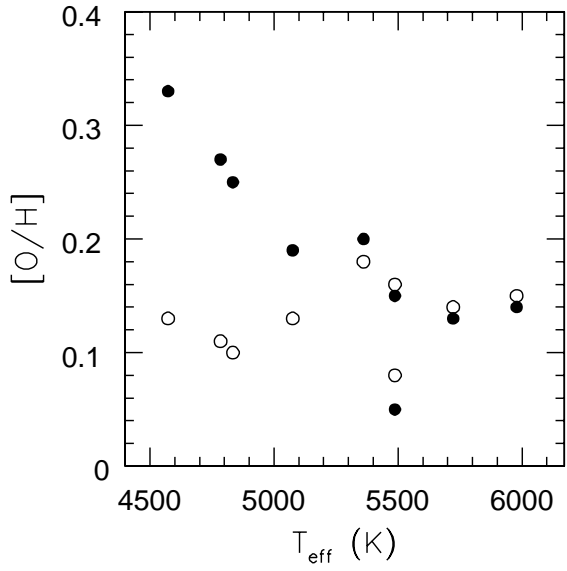}
\caption{Relative LTE O abundances as a function of \tef.  Closed circles are
the original abundances, as shown in Figure 5.  Oxygen abundances derived
without including CO in molecular equilibrium calculations- effectively
simulating the complete dissociation of the CO molecule- are given as open 
circles.}
\end{figure}

\begin{figure}
\plotone{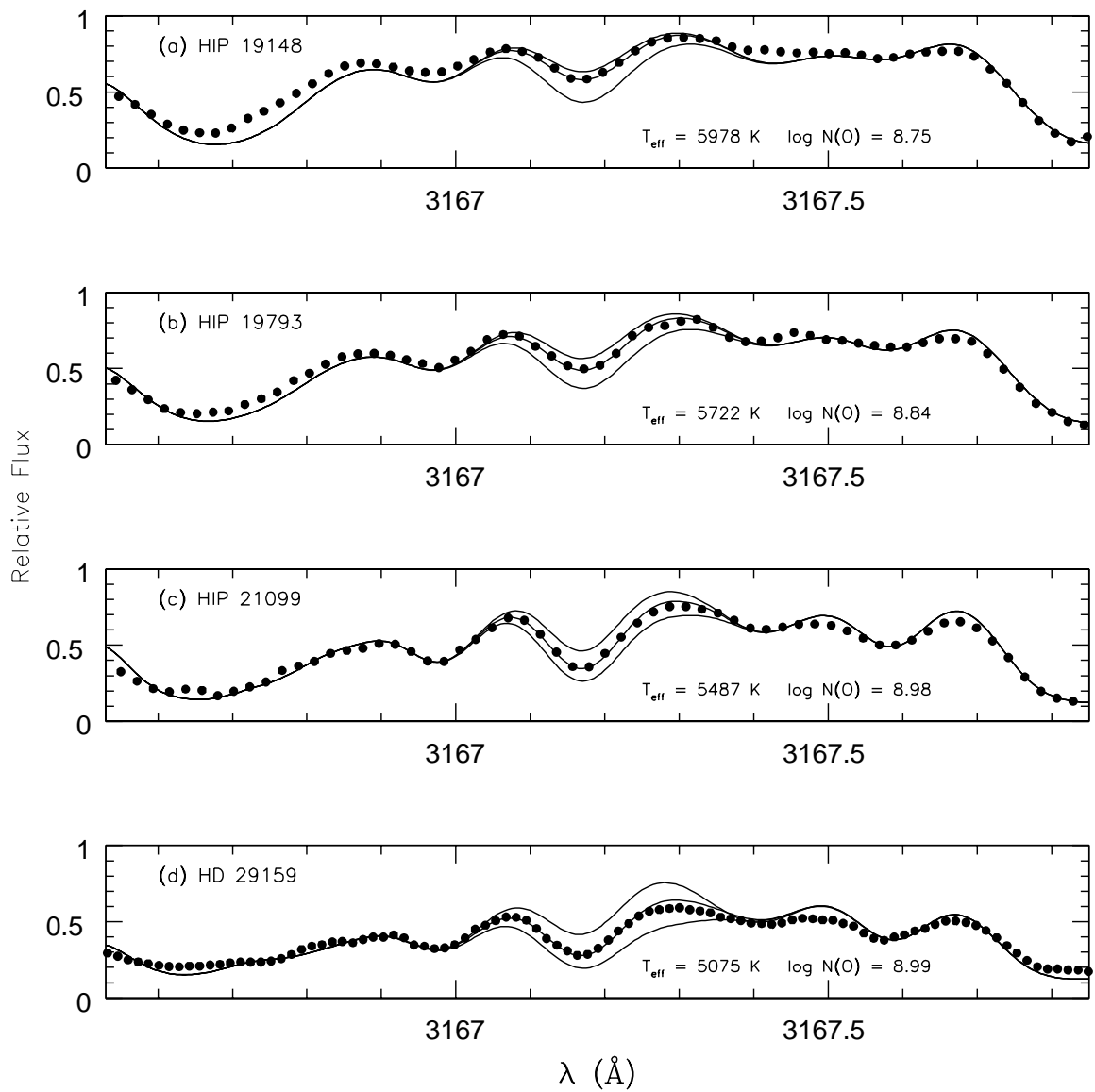}
\caption{Keck/HIRES spectra (closed circles) and synthetic fits (solid lines) of
the near-UV, $\lambda 3167$ OH line for four Hyades dwarfs.  The abundance of
the best fit is given in each panel.  The low and high abundance syntheses are 
the same for each panel and are characterized by $\log N(O) = 8.46$ and 
$\log N(O) = 9.36$.  A solar O abundance of $\log N(O) = 8.69$- the abundance 
derived from the $\lambda 6300$ [\ion{O}{1}] line in our high-quality McDonald 
2.7-m spectrum of the daytime sky- has been adopted for this analysis.}
\end{figure}

\begin{figure}
\plotone{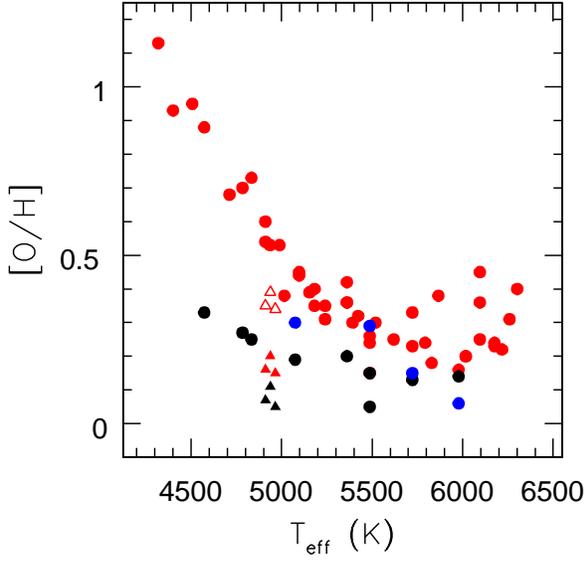}
\caption{Compilation of Hyades O abundances versus \tef.  LTE Dwarf abundances 
are given as circles and are derived from the $\lambda 3167$ OH feature (blue), 
the $\lambda 6300$ [\ion{O}{1}] line (black), and the high-excitation triplet 
(red; data taken from Schuler et al. 2005).  Giant abundances are shown as 
triangles; the abundances are derived assuming LTE from the $\lambda 6300$ 
[\ion{O}{1}] line (closed black) and the high-excitation triplet (open red).  
Giant triplet abundances subjected to the NLTE corrections of 
\citet{2003A&A...402..343T} are also given (closed red).}
\end{figure}

\begin{figure}
\plotone{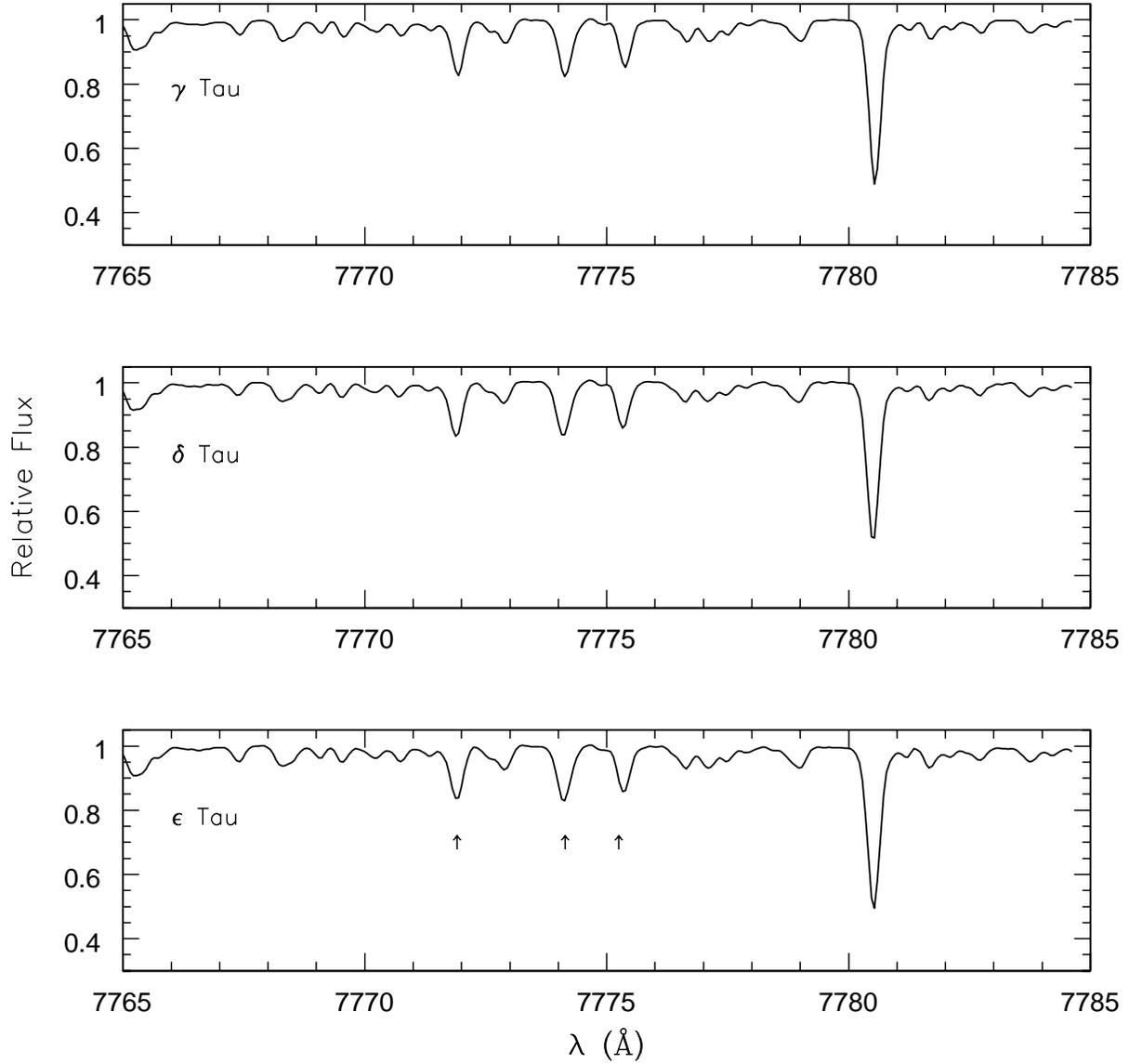}
\caption{McDonald 2.7-m/2dcoud{\'e} spectra of the high-excitation triplet 
region for the Hyades giants \gtau, \dtau, and \etau.  The triplet is marked 
by the arrows.}
\end{figure}

\begin{deluxetable}{rlrlrccccccc}
\tablecolumns{12}
\tablewidth{0pt}
\tablecaption{Cross-Identifications \& Observing Log}
\tablehead{
     \colhead{}&
     \colhead{}&
     \colhead{}&
     \colhead{}&
     \colhead{}&
     \colhead{}&
     \colhead{}&
     \colhead{}&
     \colhead{Date}&
     \colhead{Exp}&
     \colhead{Int. Time\tablenotemark{a}}&
     \colhead{}\\
     \colhead{HIP}&
     \colhead{}&
     \colhead{HD}&
     \colhead{}&
     \colhead{Other}&
     \colhead{}&
     \colhead{Telescope}&
     \colhead{}&
     \colhead{UT}&
     \colhead{\#}&
     \colhead{(s)}&
     \colhead{S/N\tablenotemark{b}}
     }

\startdata
$14976\dotfill$ && 19902 && \nodata               && McD && 12 Oct 2004 & 1 & 1800 & 225\\
$19148\dotfill$ && 25825 && $\mathrm{vB} \; 10$   && VLT && 18 Jan 2003 & 3 &  120 & 350\\
$19793\dotfill$ && 26736 && $\mathrm{vB} \; 15$   && McD && 12 Oct 2004 & 1 &  840 & 175\\
$20082\dotfill$ &&285690 && $\mathrm{vB} \; 25$   && VLT && 18 Jan 2003 & 3 &  300 & 255\\
$20949\dotfill$ &&283704 && $\mathrm{vB} \; 76$   && VLT && 18 Jan 2003 & 2 &  285 & 240\\
$21099\dotfill$ && 28593 && $\mathrm{vB} \; 87$   && VLT && 18 Jan 2003 & 3 &  220 & 330\\
$22654\dotfill$ &&284930 && \nodata               && VLT && 18 Jan 2003 & 2 &  600 & 235\\
$23312\dotfill$ &&\nodata&&$\mathrm{BD}+04 \; 810$&& VLT && 14 Oct 2002 & 2 &  340 & 240\\
\nodata         && 29159 && $\mathrm{vB} \; 99$   && VLT && 18 Jan 2003 & 3 &  400 & 340\\
                &&       &&                       &&     &&             &   &      &    \\
$20205\dotfill$ && 27371 && $\gamma \; \mathrm{Tau}$   && McD && Multiple\tablenotemark{c} & \nodata & \nodata & \nodata\\
$20455\dotfill$ && 27697 && $\delta \; \mathrm{Tau}$   && McD && Multiple & \nodata & \nodata & \nodata\\
$20889\dotfill$ && 28305 && $\epsilon \; \mathrm{Tau}$ && McD && Multiple & \nodata & \nodata & \nodata\\
                &&       &&                       &&     &&             &   &      &    \\
$\mathrm{Sun}\dotfill$ && \nodata && \nodata      && McD && 10 Oct 2004 & 1 &   30 & 950\\
\enddata

\tablenotetext{a}{Integration times for the dwarfs are the same for each exposure.}
\tablenotetext{b}{For dwarfs with more than one exposure, the S/N ratio in $\lambda 6300$ region of the coadded spectrum is given.}
\tablenotetext{c}{The specifics of each observation are given in the text.}

\end{deluxetable}

\begin{deluxetable}{lrcrrrccc}
\tablecolumns{9}
\tablewidth{0pt}
\tablecaption{Stellar Parameters}
\tablehead{
     \colhead{}&
     \colhead{}&
     \colhead{$B-V$}&
     \colhead{}&
     \colhead{$T_{\mathrm{eff}}$}&
     \colhead{}&
     \colhead{$\log g$}&
     \colhead{}&
     \colhead{$\xi$}\\
     \colhead{Star}&
     \colhead{}&
     \colhead{Mag.}&
     \colhead{}&
     \colhead{(K)}&
     \colhead{}&
     \colhead{(cgs)}&
     \colhead{}&
     \colhead{(km $\mathrm{s}^{-1}$)}
     }

\startdata
$\mathrm{HIP} \; 14976$ && 0.73 && 5487 && 4.54 && 1.24\\
$\mathrm{HIP} \; 19148$ && 0.59 && 5978 && 4.44 && 1.45\\
$\mathrm{HIP} \; 19793$ && 0.66 && 5722 && 4.49 && 1.34\\
$\mathrm{HIP} \; 20082$ && 0.98 && 4784 && 4.65 && 1.00\\
$\mathrm{HIP} \; 20949$ && 0.77 && 5361 && 4.56 && 1.20\\
$\mathrm{HIP} \; 21099$ && 0.73 && 5487 && 4.54 && 1.24\\
$\mathrm{HIP} \; 22654$ && 1.07 && 4573 && 4.68 && 1.00\\
$\mathrm{HIP} \; 23312$ && 0.96 && 4834 && 4.64 && 1.02\\
$\mathrm{HD} \; 29159\dotfill$  && 0.87 && 5075 && 4.61 && 1.09\\
                        &&      &&      &&      &&     \\
$\gamma \; \mathrm{Tau}\dotfill$   && 0.99 && 4965 && 2.63 && 1.32\\
$\delta \; \mathrm{Tau}\dotfill$   && 0.98 && 4938 && 2.69 && 1.40\\
$\epsilon \; \mathrm{Tau}\dotfill$ && 1.01 && 4911 && 2.57 && 1.47\\
                                   &&      &&      && 	   &&	  \\
$\mathrm{Sun}\dotfill$             &&      && 5777 && 4.44 && 1.38\\
\enddata

\end{deluxetable}

\begin{deluxetable}{lcccrcccccccccccccc}
\tablecolumns{19}
\tabletypesize{\small}
\rotate
\tablenum{3}
\tablewidth{0pt}
\tablecaption{Fe \& Ni Equivalent Widths- Dwarfs}
\tablehead{
     \colhead{}&
     \colhead{}&
     \colhead{}&
     \colhead{}&
     \colhead{}&
     \colhead{}&
     \colhead{}&
     \multicolumn{12}{c}{Equivalent Widths ($\mathrm{m\AA}$)}\\
     \cline{8-19}\\
     \colhead{}&
     \colhead{$\lambda_{\mathrm{rest}}$}&
     \colhead{}&
     \colhead{$\chi$}&
     \colhead{}&
     \colhead{}&
     \colhead{}&
     \colhead{}&
     \colhead{}&
     \multicolumn{8}{c}{HIP}&
     \colhead{}&
     \colhead{HD}\\
     \cline{10-17} \cline{19-19}\\
     \colhead{Ion}&
     \colhead{({\AA})}&
     \colhead{}&
     \colhead{(eV)}&
     \colhead{}&
     \colhead{$\log gf$}&
     \colhead{}&
     \colhead{$\mathrm{EW}_{\sun}$}&
     \colhead{}&
     \colhead{14976}&
     \colhead{19148}&
     \colhead{19793}&
     \colhead{20082}&
     \colhead{20949}&
     \colhead{21099}&
     \colhead{22654}&
     \colhead{23312}&
     \colhead{}&
     \colhead{29159}
     }
\startdata
\ion{Fe}{1}\ldots & 5807.79 && 3.29 && -3.41 &&  9.9 && 15.7 &\nodata& 14.8  & 23.3  & 17.5  & 15.6  & 23.7  & 22.1 && 20.0\\
		  & 5853.16 && 1.49 && -5.28 &&  8.9 && 15.4 & 7.33  & 10.7  & 29.5  & 18.2  & 17.0  & 36.1  & 28.4 && 24.5\\
		  & 6054.08 && 4.37 && -2.31 && 10.9 && 16.4 & 10.4  & 18.5  & 19.4  & 16.5  & 16.8  & 18.0  & 19.3 && 18.4\\
		  & 6120.25 && 0.91 && -5.95 && 6.53 && 12.3 &\nodata&  9.7  & 25.5  & 13.0  & 11.7  & 34.7  & 24.6 && 19.7\\
		  & 6159.38 && 4.61 && -1.97 && 14.2 && 21.7 & 15.4  & 17.7  & 26.0  & 21.6  & 21.8  & 27.6  & 25.1 && 25.2\\
		  & 6290.53 && 2.59 && -4.33 && 6.26 && 15.2 &\nodata&  8.3  & 17.2  & 10.6  & 10.4  & 17.4  & 16.5 && 14.5\\
		  & 6385.73 && 4.73 && -1.91 && 12.5 && 16.9 &\nodata& 16.6  & 22.9  & 20.0  & 17.9  & 24.7  & 22.3 && 20.0\\
		  & 6591.33 && 4.59 && -2.07 && 11.4 && 17.4 & 10.9  & 13.7  & 19.5  & 17.3  & 18.4  & 18.1  & 21.2 && 20.0\\
		  & 6608.04 && 2.28 && -4.03 && 19.4 && 28.5 & 17.2  & 25.1  & 44.4  & 33.3  & 31.6  & 44.8  & 43.6 && 39.6\\
		  & 6646.97 && 2.61 && -3.99 && 10.6 && 18.3 &\nodata& 16.9  & 29.1  & 19.7  & 20.3  & 30.1  & 30.4 && 26.0\\
		  & 6739.52 && 1.56 && -4.95 && 12.9 && 21.2 &\nodata& 15.8  & 38.9  &\nodata& 22.9  & 42.1  & 34.6 && 31.1\\
		  & 6746.98 && 2.61 && -4.35 && 5.37 &&  9.4 &\nodata&  7.7  & 16.7  & 10.3  &\nodata& 17.2  & 13.6 && 12.3\\
		  & 6837.01 && 4.59 && -1.81 && 19.2 && 26.6 & 18.7  & 25.2  & 29.0  & 25.6  & 26.7  & 26.4  & 28.2 && 27.7\\
\ion{Fe}{2}\ldots & 5425.25 && 3.20 && -3.21 && 41.6 && 44.3 & 52.4  & 51.4  &\nodata  & 40.6  & 42.1  & 31.9  &\nodata&& 33.6\\
                  & 6084.10 && 3.20 && -3.80 && 21.5 && 22.9 & 29.4  & 28.7  & 12.2  & 22.8  & 21.6  &  8.4  &\nodata&& 15.0\\
                  & 6149.25 && 3.89 && -2.72 && 40.1 && 35.9 & 50.1  & 46.2  & 20.1  & 36.0  & 37.2  & 18.8  & 22.0 && 26.4\\
		  & 6247.56 && 3.89 && -2.31 && 59.2 && 56.1 & 69.7  & 67.4  & 27.4  & 55.1  & 53.5  & 21.0  & 33.0 && 38.9\\
                  & 6432.68 && 2.89 && -3.58 && 43.0 &&\nodata& 54.6  &\nodata& 25.7  & 39.0  & 41.2  & 22.3  & 25.8 && 31.7\\
                  & 6456.39 && 3.90 && -2.08 && 66.1 && 68.5 & 82.3  & 81.4  & 36.0  & 61.4  & 65.1  & 26.2  & 41.1 && 48.1\\
\ion{Ni}{1}\ldots & 6130.14 && 4.26 && -0.96 && 22.0 && 31.6 & 25.0  & 28.6  & 29.0  & 30.7  & 30.2  & 28.1  & 34.3 && 31.6\\
		  & 6133.98 && 4.09 && -1.83 && 7.21 &&  9.5 &\nodata&\nodata& 11.5  &\nodata&\nodata&\nodata& 10.4 &&  8.8\\
		  & 6177.25 && 1.83 && -3.50 && 17.7 && 24.5 & 15.6  & 22.3  & 35.1  & 25.6  & 25.5  & 34.4  & 33.3 && 31.0\\
		  & 6223.99 && 4.10 && -0.91 && 28.4 && 39.4 & 29.2  & 38.7  & 36.7  & 37.0  & 37.6  & 34.0  & 36.8 && 38.4\\
		  & 6370.36 && 3.54 && -1.94 && 15.4 && 22.8 &\nodata& 20.5  &\nodata& 23.2  & 24.1  & 22.2  & 24.2 && 24.5\\
		  & 6414.59 && 4.15 && -1.18 && 19.3 && 26.3 & 20.4  &\nodata& 25.2  & 27.2  & 25.5  & 22.9  & 28.1 && 26.7\\
		  & 6598.61 && 4.23 && -0.98 && 26.9 && 35.3 & 28.5  & 34.3  & 35.4  & 34.3  & 36.2  & 30.7  & 34.8 && 34.4\\
		  & 6635.14 && 4.42 && -0.82 && 25.9 && 37.5 & 25.3  & 34.1  & 33.8  & 36.1  & 37.0  & 30.7  & 34.3 && 34.9\\
		  & 6842.04 && 3.66 && -1.48 && 28.3 && 39.3 & 27.6  & 37.4  & 42.4  & 38.1  & 37.9  & 41.3  & 39.9 && 41.8\\
\enddata
\end{deluxetable}

\begin{deluxetable}{lcccrccccccccc}
\tablecolumns{14}
\tabletypesize{\small}
\tablenum{4}
\tablewidth{0pt}
\tablecaption{Fe \& Ni Equivalent Widths- Giants}
\tablehead{
     \colhead{}&
     \colhead{$\lambda_{\mathrm{rest}}$}&
     \colhead{}&
     \colhead{$\xi$}&
     \colhead{}&
     \colhead{}&
     \colhead{}&
     \colhead{}&
     \colhead{}&
     \multicolumn{5}{c}{Equivalent Widths ($\mathrm{m\AA}$)}\\
     \cline{10-14}\\
     \colhead{Ion}&
     \colhead{$\mathrm{\AA}$}&
     \colhead{}&
     \colhead{(eV)}&
     \colhead{}&
     \colhead{$\log gf$}&
     \colhead{}&
     \colhead{$\mathrm{EW}_{\sun}$}&
     \colhead{}&
     \colhead{\gtau}&
     \colhead{}&
     \colhead{\dtau}&
     \colhead{}&
     \colhead{\etau}
     }

\startdata
\ion{Fe}{1}\ldots & 5853.16 && 1.49 && -5.28 &&  8.9 && 47.9 && 51.4 && 55.9  \\
		  & 6054.08 && 4.37 && -2.31 && 10.9 && 32.7 && 36.1 && 37.2  \\
		  & 6120.25 && 0.91 && -5.95 && 6.53 && 44.8 && 44.8 && 54.7  \\
		  & 6159.38 && 4.61 && -1.97 && 14.2 && 40.0 && 37.7 &&\nodata\\
		  & 6290.53 && 2.59 && -4.33 && 6.26 && 35.8 && 37.1 && 39.1  \\
		  & 6385.73 && 4.73 && -1.91 && 12.5 && 33.1 && 35.8 && 40.8  \\
		  & 6591.33 && 4.59 && -2.07 && 11.4 && 30.7 && 35.5 && 42.3  \\
		  & 6608.04 && 2.28 && -4.03 && 19.4 && 66.7 && 72.4 && 76.3  \\
		  & 6646.97 && 2.61 && -3.99 && 10.6 && 59.0 && 59.4 && 68.5  \\
		  & 6739.52 && 1.56 && -4.95 && 12.9 && 64.1 && 62.9 && 72.4  \\
		  & 6746.98 && 2.61 && -4.35 && 5.37 && 32.5 && 31.4 && 38.8  \\
		  & 6837.01 && 4.59 && -1.81 && 19.2 && 49.5 && 50.2 && 55.4  \\
\ion{Fe}{2}\ldots & 5425.25 && 3.20 && -3.21 && 41.6 && 67.4 && 67.0 && 69.8  \\ 
                  & 6084.10 && 3.20 && -3.80 && 21.5 && 47.6 && 48.0 && 48.9  \\ 
                  & 6149.25 && 3.89 && -2.72 && 40.1 && 55.3 && 54.9 && 58.5  \\
		  & 6247.56 && 3.89 && -2.31 && 59.2 && 77.3 && 76.6 && 79.7  \\
                  & 6456.39 && 3.90 && -2.08 && 66.1 && 89.2 && 90.3 && 91.1  \\     
\ion{Ni}{1}\ldots & 6130.14 && 4.26 && -0.96 && 22.0 && 49.8 && 52.6 && 58.8  \\
		  & 6133.98 && 4.09 && -1.83 && 7.21 && 26.2 && 23.1 && 23.5  \\
		  & 6223.99 && 4.10 && -0.91 && 28.4 && 66.3 && 61.3 &&\nodata\\
		  & 6598.61 && 4.23 && -0.98 && 26.9 && 55.6 && 56.2 && 64.1  \\
		  & 6635.14 && 4.42 && -0.82 && 25.9 && 63.9 && 60.2 && 66.9  \\
		  & 6842.04 && 3.66 && -1.48 && 28.3 && 68.4 && 70.5 && 71.3  \\

\enddata

\end{deluxetable}

\begin{deluxetable}{lrccccccccccccccccc}
\tablecolumns{19}
\rotate
\tabletypesize{\small}
\tablenum{5}
\tablewidth{0pt}
\tablecaption{Final Fe and Ni Abundances}
\tablehead{
     \colhead{}&
     \colhead{}&
     \multicolumn{5}{c}{[\ion{Fe}{1}/H]}&
     \colhead{}&
     \multicolumn{5}{c}{[\ion{Fe}{2}/H]}&
     \colhead{}&
     \multicolumn{5}{c}{[\ion{Ni}{1}/H]}\\
     \cline{3-7} \cline{9-13} \cline{15-19}\\
     \colhead{Star}&
     \colhead{}&
     \colhead{OVER}&
     \colhead{$\sigma$}&
     \colhead{}&
     \colhead{MLT5}&
     \colhead{$\sigma$}&
     \colhead{}&
     \colhead{OVER}&
     \colhead{$\sigma$}&
     \colhead{}&
     \colhead{MLT5}&
     \colhead{$\sigma$}&
     \colhead{}&
     \colhead{OVER}&
     \colhead{$\sigma$}&
     \colhead{}&
     \colhead{MLT5}&
     \colhead{$\sigma$}
     }
\startdata
$\mathrm{HIP} \; 14976$            && 0.07 & 0.03 && 0.08 & 0.03 && 0.26 & 0.05 && 0.27 & 0.05 && 0.13 & 0.03 && 0.13 & 0.03\\
$\mathrm{HIP} \; 19148$            && 0.10 & 0.04 && 0.09 & 0.03 && 0.18 & 0.07 && 0.18 & 0.07 && 0.11 & 0.03 && 0.10 & 0.03\\
$\mathrm{HIP} \; 19793$            && 0.13 & 0.03 && 0.13 & 0.03 && 0.29 & 0.05 && 0.30 & 0.05 && 0.16 & 0.03 && 0.16 & 0.03\\
$\mathrm{HIP} \; 20082$            && 0.09 & 0.03 && 0.10 & 0.03 && 0.52 & 0.11 && 0.49 & 0.12 && 0.22 & 0.04 && 0.22 & 0.04\\
$\mathrm{HIP} \; 20949$            && 0.04 & 0.03 && 0.05 & 0.03 && 0.32 & 0.07 && 0.32 & 0.09 && 0.11 & 0.04 && 0.12 & 0.04\\
$\mathrm{HIP} \; 21099$            && 0.08 & 0.04 && 0.09 & 0.04 && 0.22 & 0.08 && 0.23 & 0.07 && 0.13 & 0.04 && 0.14 & 0.04\\
$\mathrm{HIP} \; 22654$            && 0.14 & 0.04 && 0.15 & 0.04 && 0.74 & 0.14 && 0.69 & 0.13 && 0.25 & 0.05 && 0.24 & 0.04\\
$\mathrm{HIP} \; 23312$            && 0.07 & 0.03 && 0.08 & 0.03 && 0.53 & 0.09 && 0.51 & 0.10 && 0.20 & 0.03 && 0.21 & 0.04\\
$\mathrm{HD} \; 29159\dotfill$     && 0.04 & 0.03 && 0.05 & 0.03 && 0.37 & 0.09 && 0.35 & 0.09 && 0.14 & 0.04 && 0.14 & 0.04\\
                                   &&	   &	  &&	  &	 &&	 &	&&	&      &&      &      &&      &     \\
$\gamma \; \mathrm{Tau}\dotfill$   && 0.14 & 0.08 && 0.15 & 0.08 && 0.22 & 0.16 && 0.23 & 0.15 && 0.26 & 0.07 && 0.27 & 0.07\\
$\delta \; \mathrm{Tau}\dotfill$   && 0.14 & 0.07 && 0.15 & 0.08 && 0.26 & 0.16 && 0.27 & 0.16 && 0.23 & 0.06 && 0.25 & 0.06\\
$\epsilon \; \mathrm{Tau}\dotfill$ && 0.20 & 0.08 && 0.21 & 0.08 && 0.22 & 0.16 && 0.22 & 0.15 && 0.26 & 0.08 && 0.27 & 0.08\\
\enddata

\end{deluxetable}

\begin{deluxetable}{lrccccccc}
\tablecolumns{9}
\tablewidth{0pt}
\tablenum{6}
\tablecaption{[\ion{O}{1}] Equivalent Widths and Abundances}
\tablehead{
     \colhead{}&
     \colhead{}&
     \colhead{}&
     \colhead{}&
     \multicolumn{2}{c}{OVER}&
     \colhead{}&
     \multicolumn{2}{c}{MLT5}\\
     \cline{5-6} \cline{8-9}\\
     \colhead{}&
     \colhead{}&
     \colhead{EW\tablenotemark{a}}&
     \colhead{}&
     \colhead{$\log N(\mathrm{O})$}&
     \colhead{[O/H]}&
     \colhead{}&
     \colhead{$\log N(\mathrm{O})$}&
     \colhead{[O/H]}\\
     \colhead{Star}&
     \colhead{}&
     \colhead{($\mathrm{m\AA}$)}&
     \colhead{}&
     \colhead{(dex)}&
     \colhead{(dex)}&
     \colhead{}&
     \colhead{(dex)}&
     \colhead{(dex)}
     }

\startdata
$\mathrm{Sun}\dotfill$             && 5.5 && 8.69 &\nodata  && 8.66 &\nodata \\
                                   &&     &&      &	    &&      & \\
$\mathrm{HIP} \; 14976$            && 6.6 && 8.74 & 0.19 && 8.71 & 0.05\\
$\mathrm{HIP} \; 19148$            && 6.1 && 8.83 & 0.05 && 8.80 & 0.14\\
$\mathrm{HIP} \; 19793$            && 6.9 && 8.82 & 0.14 && 8.79 & 0.13\\
$\mathrm{HIP} \; 20082$            && 7.7 && 8.96 & 0.13 && 8.94 & 0.28\\
$\mathrm{HIP} \; 20949$            && 7.6 && 8.89 & 0.27 && 8.86 & 0.20\\
$\mathrm{HIP} \; 21099$            && 7.3 && 8.84 & 0.20 && 8.81 & 0.15\\
$\mathrm{HIP} \; 22654$            && 8.0 && 9.02 & 0.15 && 9.00 & 0.34\\
$\mathrm{HIP} \; 23312$            && 7.5 && 8.94 & 0.33 && 8.90 & 0.24\\
$\mathrm{HD} \; 29159\dotfill$     && 7.4 && 8.88 & 0.19 && 8.85 & 0.19\\
                                   &&     &&      &	 &&	 &     \\
$\gamma \; \mathrm{Tau}\dotfill$   && 28.6 && 8.74 & 0.05 && 8.72 & 0.06\\
$\delta \; \mathrm{Tau}\dotfill$   && 29.3 && 8.80 & 0.11 && 8.79 & 0.13\\
$\epsilon \; \mathrm{Tau}\dotfill$ && 31.0 && 8.76 & 0.07 && 8.75 & 0.09\\
\enddata

\tablenotetext{a}{Equivalent widths are not corrected for the \ion{Ni}{1} blend.}

\end{deluxetable}

\begin{deluxetable}{lccccccccccclcc}
\tablecolumns{15}
\tablewidth{0pt}
\tablenum{7}
\tabletypesize{\footnotesize}
\tablecaption{O Abundance Uncertainties}
\tablehead{
     \colhead{Star}&
     \colhead{}&
     \colhead{$T_{\mathrm{eff}}$}&
     \colhead{}&
     \colhead{$\log g$}&
     \colhead{}&
     \colhead{$\xi$}&
     \colhead{}&
     \colhead{EW}&
     \colhead{}&
     \colhead{Ni}&
     \colhead{}&
     \colhead{C}&
     \colhead{}&
     \colhead{$\sigma_{\mathrm{Total}}$}
     }

\startdata
$\mathrm{HIP} \; 14976$ && 0.00 && 0.06 && 0.00 && 0.14 && $-0.06/0.05$ && $0.04/-0.02$ && $\pm0.18$\\
$\mathrm{HIP} \; 19148$ && 0.02 && 0.04 && 0.00 && 0.10 && $-0.05/0.05$ && $0.01/0.00$  && $\pm0.12$\\
$\mathrm{HIP} \; 19793$ && 0.01 && 0.05 && 0.00 && 0.11 && $-0.05/0.05$ && $0.02/-0.02$ && $\pm0.15$\\
$\mathrm{HIP} \; 20082$ && 0.00 && 0.04 && 0.00 && 0.08 && $-0.03/0.03$ && $0.07/-0.05$ && $\pm0.13$\\
$\mathrm{HIP} \; 20949$ && 0.00 && 0.05 && 0.00 && 0.11 && $-0.06/0.04$ && $0.04/-0.03$ && $\pm0.15$\\
$\mathrm{HIP} \; 21099$ && 0.00 && 0.06 && 0.00 && 0.11 && $-0.06/0.05$ && $0.04/-0.02$ && $\pm0.16$\\
$\mathrm{HIP} \; 22654$ && 0.00 && 0.03 && 0.00 && 0.05 && $-0.02/0.03$ && $0.07/-0.05$ && $\pm0.11$\\
$\mathrm{HIP} \; 23312$ && 0.00 && 0.04 && 0.00 && 0.08 && $-0.04/0.03$ && $0.07/-0.05$ && $\pm0.13$\\
$\mathrm{HD} \; 29159\dotfill$ && 0.00 && 0.05 && 0.00 && 0.12 && $-0.05/0.04$ && $0.06/-0.04$ && $\pm0.16$\\
    
                                   &&      && 	   &&      && 	   &&   	   &&		   &&          \\
$\gamma \; \mathrm{Tau}\dotfill$   && 0.00 && 0.08 && 0.00 && 0.02 && $-0.04/0.03$ && $0.03/-0.02$ && $\pm0.10$\\
$\delta \; \mathrm{Tau}\dotfill$   && 0.00 && 0.08 && 0.00 && 0.02 && $-0.03/0.02$ && $0.03/-0.03$ && $\pm0.09$\\
$\epsilon \; \mathrm{Tau}\dotfill$ && 0.00 && 0.08 && 0.00 && 0.02 && $-0.03/0.02$ && $0.03/-0.02$ && $\pm0.09$\\
\enddata

\end{deluxetable}

\begin{deluxetable}{rccccccccccccccc}
\tablecolumns{15}
\tabletypesize{\small}
\rotate
\tablenum{8}
\tablewidth{0pt}
\tablecaption{Additional Dwarf [Ni/H] Abundances}
\tablehead{
     \colhead{$\lambda_{\mathrm{rest}}$}&
     \colhead{}&
     \colhead{$\chi$}&
     \colhead{}&
     \colhead{}&
     \multicolumn{8}{c}{HIP}&
     \colhead{}&
     \colhead{HD}\\
     \cline{6-13} \cline{15-15}\\
     \colhead{({\AA})}&
     \colhead{}&
     \colhead{(eV)}&
     \colhead{}&
     \colhead{$\log N_{\odot}$}&
     \colhead{14976}&
     \colhead{19148}&
     \colhead{19793}&
     \colhead{20082}&
     \colhead{20949}&
     \colhead{21099}&
     \colhead{22654}&
     \colhead{23312}&
     \colhead{}&
     \colhead{29159}
     }
\startdata
5996.73 && 4.24 && 6.34 & 0.07 &\nodata& 0.16 &\nodata&\nodata&\nodata&\nodata&\nodata &&\nodata\\
6025.75 && 4.24 && 6.32 & 0.15 &\nodata& 0.13 & 0.25  & 0.09  & 0.10  &\nodata& 0.27   && 0.19  \\
6053.68 && 4.24 && 6.36 & 0.12 & 0.06  & 0.14 & 0.17  & 0.07  & 0.07  & 0.22  & 0.19   && 0.13  \\
6130.14 && 4.26 && 6.29 & 0.14 & 0.12  & 0.13 & 0.22  & 0.09  & 0.09  & 0.25  & 0.26   && 0.15  \\
6598.61 && 4.23 && 6.33 & 0.11 & 0.13  & 0.16 & 0.25  & 0.10  &\nodata& 0.25  & 0.21   && 0.14  \\
        &&      &&      &      &       &      &       &       &       &       &        &&       \\
Mean    &&      &&      & 0.12 & 0.10  & 0.14 & 0.22  & 0.09  & 0.09  & 0.24  & 0.23   && 0.15  \\
$\sigma_{mean}$&& &&    & 0.02 & 0.03  & 0.01 & 0.02  & 0.01  & 0.01  & 0.01  & 0.02   && 0.02  \\

\enddata
\end{deluxetable}

\begin{deluxetable}{lcccccccccccc}
\tablecolumns{13}
\tabletypesize{\small}
\tablenum{9}
\rotate
\tablewidth{0pt}
\tablecaption{\ion{O}{1} Triplet- Giants}
\tablehead{
     \colhead{}&
     \colhead{}&
     \colhead{$\mathrm{EW}_{7772}$}&
     \colhead{$\mathrm{EW}_{7774}$}&
     \colhead{$\mathrm{EW}_{7775}$}&
     \colhead{}&
     \multicolumn{3}{c}{LTE}&
     \colhead{}&
     \multicolumn{3}{c}{NLTE}\\
     \cline{7-9} \cline{11-13}\\
     \colhead{Star}&
     \colhead{}&
     \colhead{(m$\mathrm{\AA}$)}&
     \colhead{(m$\mathrm{\AA}$)}&
     \colhead{(m$\mathrm{\AA}$)}&
     \colhead{}&
     \colhead{$\mathrm{[O/H]}_{7772}$\tablenotemark{a}}&    
     \colhead{$\mathrm{[O/H]}_{7774}$}&    
     \colhead{$\mathrm{[O/H]}_{7775}$}&    
     \colhead{}&    
     \colhead{$\mathrm{[O/H]}_{7772}$}&    
     \colhead{$\mathrm{[O/H]}_{7774}$}&    
     \colhead{$\mathrm{[O/H]}_{7775}$}
     }
\startdata
\gtau && 58.8 & 61.4 & 44.9 && 0.35 & 0.52 & 0.33 && 0.14 & 0.31 & 0.16\\
\dtau && 56.7 & 60.1 & 44.6 && 0.38 & 0.57 & 0.40 && 0.17 & 0.36 & 0.23\\
\etau && 56.2 & 61.4 & 45.1 && 0.33 & 0.55 & 0.37 && 0.12 & 0.34 & 0.20\\
\enddata

\tablenotetext{a}{Solar triplet-based O abundances are from Schuler et al. 
2005.}

\end{deluxetable}

\end{document}